\documentclass[11pt,reqno]{article}
\usepackage{amsmath}
\usepackage{amsfonts}
\usepackage{amssymb}
\usepackage{latexsym}
\usepackage[dvips]{graphicx}
\usepackage{epsf}
\usepackage{color}
\usepackage{bbm}
\usepackage{ulem}

\allowdisplaybreaks \numberwithin{equation}{section}

\newcommand{\half}{\tfrac{1}{2}}
\newcommand{\p}{\partial}

\newcommand{\be}{\begin{equation}}
\newcommand{\ee}{\end{equation}}
\newcommand{\bi}{\begin{itemize}}
\newcommand{\ei}{\end{itemize}}
\newcommand{\bea}{\begin{eqnarray}}
\newcommand{\eea}{\end{eqnarray}}
\newcommand{\nn}{\nonumber}

\let\a=\alpha \let\b=\beta  \let\g=\gamma  \let\d=\delta
\let\z=\zeta        
\let\m=\mu    \let\n=\nu          
\let\s=\sigma \let\t=\tau    \let\ph=\phi 
    
\let\G=\Gamma     
    \let\Si=\Sigma     
  \let\eps=\epsilon

\newcommand{\Db}{\bar{D}}
\newcommand{\Rb}{\bar{R}}

\newcommand{\cD}{\mathcal{D}}

\newcommand{\cL}{\mathcal{L}}
\newcommand{\cM}{\mathcal{M}}
\newcommand{\cN}{\mathcal{N}}
\newcommand{\cO}{\mathcal{O}}
\newcommand{\cR}{\mathcal{R}}

\newcommand{\cT}{\mathcal{T}}

\newcommand{\cZ}{\mathcal{Z}}

\newcommand{\tN}{\tilde{N}}
\newcommand{\ts}{\tilde{\sigma}}

\newcommand{\hN}{\hat{N}}
\newcommand{\hs}{\hat{\sigma}}

\newcommand{\bN}{\bar{N}}
\newcommand{\bs}{\bar{\sigma}}

\begin{document}


\thispagestyle{empty}
\begin{flushright} \small
MZ-TH/12-56
\end{flushright}
\bigskip

\begin{center}
 {\LARGE\bfseries  A functional renormalization group equation} \\[1.5ex] 
{\LARGE\bfseries   for foliated spacetimes}  \\[2cm]

{\Large Stefan Rechenberger and Frank Saueressig} \\[1.5ex]
{\it PRISMA Cluster of Excellence \& Institute for Physics (THEP)}\\
{\it University of Mainz, Staudingerweg 7, D-55099 Mainz, Germany}

\end{center}
\vspace{20mm}

\hrule\bigskip

\centerline{\bfseries Abstract} \medskip
\noindent
We derive an exact functional renormalization group equation for the projectable version of Ho\v{r}ava-Lifshitz gravity. The flow equation encodes the gravitational degrees of freedom in terms of the lapse function, shift vector and spatial metric and is manifestly invariant under background foliation-preserving diffeomorphisms. Its relation to similar flow equations for gravity in the metric formalism is discussed in detail, and we argue that the space of action functionals, invariant under the full diffeomorphism group, forms a subspace of the latter invariant under renormalization group transformations. As a first application we study the RG flow of the Newton constant and the cosmological constant in the ADM formalism.  In particular we show that the non-Gaussian fixed point found in the metric formulation is qualitatively unaffected by the change of variables and persists also for Lorentzian signature metrics.
\bigskip
\hrule\bigskip
\newpage

\section{Introduction}
The construction of a consistent and predictive quantum theory for gravity constitutes one of the major challenges for theoretical high energy physics to date. Within the realm of quantum field theory there are currently two proposals that receive a lot of attention. Quantum Einstein Gravity (QEG) \cite{Niedermaier:2006wt,Reuter:2012id,Percacci:2011fr,Litim:2011cp} is based on Weinberg's Asymptotic Safety scenario \cite{wein,Weinproc1} and suggests that gravity could be a non-perturbatively renormalizable quantum field theory. In this case the UV-completion of the theory is provided by a non-Gaussian fixed point (NGFP) of the renormalization group (RG) flow. For RG trajectories that are captured by the NGFP in the UV, this construction ensures that physical quantities are free from unphysical UV divergences. Provided that the UV fixed point comes with a finite number of relevant directions, this construction is predictive. Starting from the seminal work \cite{Reuter:1996cp} there is now a solid body of evidence supporting the existence and predictivity of this NGFP \cite{Dou:1997fg,Lauscher:2002sq,Fischer:2006fz,Codello:2007bd,Codello:2008vh,Benedetti:2009gn,Benedetti:2010nr,Groh:2010ta,Eichhorn:2009ah,Eichhorn:2010tb}. The more recent proposal advocated by Ho\v{r}ava \cite{Horava:2008ih} suggests that the UV completion of gravity is an action which is anisotropic in the space and time coordinates. Higher powers of the spatial momenta entering into the gravitational propagators could render this construction perturbatively renormalizable \cite{Orlando:2009en,Giribet:2010th} but lead to difficulties when trying to restore Lorentz invariance in the low energy regime, see \cite{Horava:2011gd,Visser:2011mf,Briscese:2012rz} for selected reviews and further references.

A crucial ingredient in investigating both scenarios is the RG flow of the theory. For practical computations, this flow can conveniently be encoded by the Wetterich equation \cite{Wetterich:1992yh} which captures the scale dependence of the effective average action $\Gamma_k$ and can schematically be written as
\be\label{FRGE}
k \p_k \Gamma_k[\phi, \bar{\phi}] = \half {\rm Tr} \left[ \left( \Gamma_k^{(2)} + \cR_k \right)^{-1} k \p_k \cR_k \right] \, . 
\ee
Here, $\Gamma_k^{(2)}$ denotes the second functional derivative of the effective average action with respect to the fluctuation fields $\phi$ at fixed background $\bar{\phi}$ and {\rm Tr} contains a sum over all fields of the theory and an integration over loop-momenta. Furthermore, $\cR_k$ is a matrix-valued IR cutoff, which provides a $k$-dependent mass term for fluctuations with momenta $p^2<k^2$. The interplay between the full regularized propagator $\left( \Gamma_k^{(2)}  + \cR_k \right)^{-1}$ and $k \p_k \cR_k$ ensures that the Tr receives contributions from a small $p^2$-interval around $k^2 \approx p^2$ only, rendering the trace contribution finite.

The flow equation \eqref{FRGE} is defined on the so-called theory space. This space contains all action functionals that can be build from a given field content and are compatible with the symmetries of the theory under consideration. Considering metric QEG, for example, the theory space $\cT^{\rm mQEG}$ contains
all action functionals build from the spacetime metric $g_{\m\n}$, which are invariant under general coordinate transformations Diff($\cM$).

An interesting open question at this stage is if the metric $g_{\m\n}$ indeed provides the correct formulation for the gravitational degrees of freedom. Coupling gravity to fermionic matter degrees of freedom suggests that a first-order formalism based on the vielbein may be more fundamental. A first investigation of this scenario was initiated in \cite{Daum:2010qt} utilizing the Wetterich equation tailored to the fundamental fields of Einstein-Cartan gravity, i.e., the vielbein and the spin connection. Along a different line, the existence of time points at a preferred direction which in terms of Euclidean geometry may reflect itself in a foliation structure of spacetime. Such a structure is naturally captured by the ADM decomposition of the metric \cite{Arnowitt:1959ah}. While all these formulations may be on-shell equivalent at the classical level, it is not clear if they describe the same quantum theory. In particular it is a priori unclear if the NGFP underlying Asymptotic Safety in the metric formulation also appears when using different fields for encoding gravity and, if so, weather these descriptions  fall into the same universality class, in the sense that the universal critical exponents associated with the NGFPs actually coincide.

In this paper, we address these issues by adapting the Wetterich equation \eqref{FRGE} to the ADM decomposition, thereby implementing a foliation structure on the underlying quantum spacetime. In this case the metric $g_{\mu\nu}$ is decomposed into a lapse function $N$, a shift-vector $N_i$ and a metric on spatial slices $\sigma_{ij}$
\be\label{ADMdec1}
g_{\m\n} \mapsto \{ \, N \, , \, N_i \, , \, \sigma_{ij} \, \} \, . 
\ee
The lapse function and the shift vector essentially describe how the spatial slices are welded together, thus imprinting the (Euclidean) spacetime $\cM$ with a preferred direction. This decomposition then naturally entails a spacetime structure that is topologically $\cM = S^1 \times \Sigma$, where $\Sigma$ are the $d$-dimensional leaves of the foliation. The direction singled out by the $S^1$ allows to Wick-rotate between Euclidean and Lorentzian signature, and will thus be referred to as (Euclidean) time direction. Notably, the ADM construction is very close to the geometric setting underlying the Monte-Carlo simulations of Causal Dynamical Triangulations (CDT) \cite{ajl1,ajl5,Benedetti:2009ge,Kommu:2011wd,Ambjorn:2012jv}.

\begin{table}[t]
\begin{center}
\begin{tabular}{|c|c|c|c|} \hline
theory & theory space & gravitational fields & symmetry \\ \hline \hline
metric QEG & $\cT^{\rm mQEG}$ & $\gamma_{\m\n}({\bf x})$ & Diff($\cM$) \\ \hline
foliated QEG & $\cT^{\rm fQEG}$ & $\, N(\tau, \vec{x}) \, , \, N_i(\tau, \vec{x}) \, , \, \sigma_{ij}(\tau, \vec{x}) $ & Diff($\cM$) \\ 
projectable Ho\v{r}ava-Lifshitz & $\cT^{\rm pHL}$ & $ \, N(\tau) \, , \, N_i(\tau, \vec{x}) \, , \, \sigma_{ij}(\tau, \vec{x})$ & Diff($\cM, \Sigma$) \\
non-projectable Ho\v{r}ava-Lifshitz & $\cT^{\rm npHL}$ & $ \, N(\tau, \vec{x}) \, , \, N_i(\tau, \vec{x}) \, , \, \sigma_{ij}(\tau, \vec{x})$ & Diff($\cM, \Sigma$) \\ \hline
\end{tabular}
\end{center}
\caption{\label{Tab.theo} Definition of the gravitational theory spaces $\cT$ that emerge within the ADM decomposition of the metric field \eqref{ADMdec1}. According to their precise field content and underlying symmetry group they constitute natural generalizations of the theory space underlying metric QEG, see \eqref{embedd}.}
\end{table}
Depending on the precise field content and symmetry groups there are several theory spaces that can naturally be associated with the decomposition \eqref{ADMdec1}, see Table \ref{Tab.theo}. Firstly, one can insist that the symmetry group acting on the ADM fields is the full Diff($\cM$) symmetry. In this case eq.\ \eqref{ADMdec1} constitutes a non-linear field-redefinition of the gravitational degrees of freedom.
The resulting theory space of foliated QEG, $\cT^{\rm fQEG}$, is equivalent to the one of metric QEG, $\cT^{\rm mQEG}$. Its interaction monomials are constructed from the ADM fields and preserve Diff($\cM$). The foliation structure of $\cM$ naturally defines a subgroup of Diff($\cM$), the foliation-preserving diffeomorphisms Diff($\cM, \Sigma$), eq.\ \eqref{gaugeVariations2}. Adopting this subgroup as the symmetry group of the theory space gives rise to the theory spaces underlying Ho\v{r}ava-Lifshitz (HL) gravity \cite{Horava:2008ih}. In its projectable version (pHL), the lapse function $N(\tau)$ depends on time only.
In terms of the field content, this can be understood as a partial gauge fixing eliminating the space dependence of the lapse function living on $\cT^{\rm fQEG}$. The weaker symmetry requirements on $\cT^{\rm pHL}$ allow to write additional interaction monomials for $\Gamma_k$, which are invariant under Diff($\cM, \Sigma$) but break Diff($\cM$) invariance. Thus $\cT^{\rm fQEG}$ is embedded into the theory space of projectable HL theory $ \cT^{\rm pHL}$. Finally, the non-projectable version of HL theory also includes a spatial dependence in the lapse function. Thus $ \cT^{\rm npHL}$ contains additional interaction monomials not present in $\cT^{\rm pHL}$. Based on these different symmetry requirements the theory spaces thus satisfy
\be\label{embedd}
\cT^{\rm mQEG} = \cT^{\rm fQEG} \subset \cT^{\rm pHL} \subset \cT^{\rm npHL}\, . 
\ee

As it will turn out, the natural symmetry of the Wetterich equation formulated in terms of the ADM fields are foliation-preserving diffeomorphisms. The key  observation underlying this assessment is that Diff$(\cM)$ acts non-linearly on the ADM fields. Since it is a key requirement in the derivation of a Wetterich-type flow equation that its regulator is quadratic in the fluctuation fields, it is impossible to retain invariance with respect to a non-linearly realized symmetry. Thus our functional renormalization group equation (FRGE) is invariant under background foliation-preserving diffeomorphisms only, since Diff$(\cM, \Sigma)$ is the maximal subgroup that is realized by linear transformations. This restriction of the symmetry group has a direct consequence for the theory space on which our FRGE is formulated: besides containing all interaction monomials that are invariant under the Diff($\cM$), the theory space also contains interactions which are invariant under foliation preserving diffeomorphisms and thus break the full diffeomorphism invariance explicitly. Thus our flow equation naturally encodes the RG flow on $\cT^{\rm pHL}$.\footnote{As it will turn out, the off-shell setting of the FRGE implies that  foliated QEG and projective Ho\v{r}ava-Lifshitz gravity are actually described by the same flow equation.} 

The relation \eqref{embedd} indicates that $\cT^{\rm mQEG}$ can be embedded into $\cT^{\rm pHL}$. An important result following from \cite{Reuter:1996cp} is that the subspace $\cT^{\rm mQEG}$ is actually {\it closed under the RG flow}. If the RG flow starts from an action functional preserving full diffeomorphism invariance, $\Gamma^{\rm mQEG}_{\hat k} \in \cT^{\rm mQEG}$, integrating out quantum fluctuations will not generate interactions that violate Diff($\cM$) dynamically. This leaves the phenomenologically interesting scenario that $\cT^{\rm mQEG}$ serves as an IR-attractor within $\cT^{\rm pHL}$, leading to a dynamical restoration of Lorentz symmetry at low energies.

The foliation structure $\cM = S^1 \times \Sigma$ will actually lead to a FRGE which closely resembles
the ones encountered for quantum field theories at finite temperature. In this interpretation, the system is seen as $d$-dimensional gravity coupled to a scalar $N$ and vector $N_i$ {\it at finite temperature} $T$ proportional to the radius $R$ of the $S^1$. In this setting quantum fluctuations are split into two classes: spatial fluctuations of the $d$-dimensional system at $R=0$ (zero temperature) and thermal fluctuations associated with the circle $S^1$. The latter are captured by the Matsubara modes of the system. The FRGE 
constructed in this paper integrates out both quantum and thermal fluctuations, essentially relying on the ``imaginary time'' formalism of finite temperature quantum field theory. This analogy allows to parallel the construction of Wilsonian flow equations for thermal field theories \cite{Litim:1998yn,Litim:2006ag,Floerchinger:2011sc}, adapting it to the gravitational system at hand.

The rest of this work is organized as follows. We start by reviewing the classical ADM construction \cite{Arnowitt:1959ah} in Sect.\ \ref{sect.2}. The Wetterich equation capturing the RG flow on $\cT^{\rm pHL}$ is constructed in Sect.\ \ref{sect.3}. This flow equation respects foliation-preserving diffeomorphisms as a background symmetry, which severely restricts the interaction monomials that can be generated dynamically by the RG flow. As a first application we consider a truncation of the effective average action $\Gamma_k$ which does not contain anisotropic couplings and derive the signature-dependent beta functions of the ADM-decomposed Einstein-Hilbert truncation in Sect.\ \ref{sect:4}.\footnote{First results on the RG flow on $\cT^{\rm pHL}$ including Diff($\cM$)-breaking interactions will be reported elsewhere \cite{inprep}.} Their properties and fixed point structure, which have partially been reported previously \cite{Manrique:2011jc}, are studied in
 in Sect.\ \ref{sect.5}.  We close with a summary and an outlook on possible applications in Sect.\ \ref{sect.6}. The technical details entering the construction of the beta functions have been collected in two appendices.

\section{Decomposing spacetime into space and time}
\label{sect.2}
In order to make the present work self-contained and fix our notation, we start by reviewing the classical ADM decomposition of the metric field \cite{Arnowitt:1959ah}. This construction plays an essential role when studying gravitational RG flows on spacetimes carrying a foliation structure or exhibiting anisotropic-scaling effects between space and time.

The ADM construction starts from a $D$-dimensional Riemannian manifold $\cM$ with metric $\gamma_{\a\b}$. The spacetime signature can either be Euclidean, $(+,+,\ldots)$, or Lorentzian, $(-,+,+,\ldots)$ and we use the signature parameter $\epsilon \equiv {\rm sig} \gamma_{\a\b} = \pm 1$ to distinguish the two cases. On this manifold we introduce a time function $\t (x)$ relating a real number, which we call time $\t$, to every spacetime point $x$. This function equips $\cM$ with a vector field $\p_\a \tau$ which, for Lorentzian signature, is taken to be timelike. The $D$-dimensional spacetime manifold can then be seen as a stack of spatial slices $\Si_{\t_i}=\{ x: \t (x)=\t_i \}$ with spatial dimension $d=(D-1)$. The hypersurfaces $\Si_{\t_i}$ come with a normal vector $n_\a$, which we take as future-directed ($n^\a\p_\a\t > 0$) and normalized to unity ($\gamma^{\a\b} n_\a n_\b = \eps$). It can be related to the time function by $n_\a = \eps \tN \p_\a \t$ where the Lapse function $\tN$ acts as a normalization factor.
 
On each spatial slice we introduce coordinates $y^i$, $i = 1, \ldots, d$. On neighboring slices these coordinate systems will be related by the integral curves along $\p_\a \tau$. Explicitly we choose $y^i$ to be constant along such a curve. The Jacobian relating the coordinate systems $x^\m$ and $(\t, y^i)$ is given by
\be\label{coc}
t^\a = \left. \frac{\p x^\a}{\p \t} \right|_{y^i} \, , \qquad\qquad    e^\a_i = \left. \frac{\p x^\a}{\p y^i} \right|_\t \, .
\ee
Note that $n_\a \, e^\a_i = 0$ holds, since $n_\a$ is normal to the hypersurfaces. The vector $t^\a$ can be decomposed into its components tangential and perpendicular to the spatial slices as
\be\label{decomp}
t^\a = \tN n^\a + \tN^i e^\a_i \, .
\ee
Here, the $d$-dimensional shift vector $\tN^i$ is purely spatial. 

For the coordinate one-forms the change of coordinates \eqref{coc} implies
\be
dx^\a = t^\a d\t + e^\a_i dy^i = \tN n^\a d\t + \left( dy^i + \tN^i d\t \right)e^\a_i \, .
\ee
Consequently, the infinitesimal squared line element is given by
\be\label{lineelement}
ds^2 = \g_{\a\b} dx^\a dx^\b = \eps \tN^2 d\t^2 + \ts_{ij}\left( dy^i + \tN^i d\t \right)\left( dy^j + \tN^j d\t \right) \, ,
\ee
where $\ts_{ij} = \gamma_{\a \b} \, e^\a_i e^\b_j$ is the induced metric on the spatial slices and the component fields depend on the spacetime coordinates
\be\label{folQEG}
 \tilde{\chi}^{\rm fQEG} = \{ \, \tN(\t, x^i) \, , \, \tN_i(\t, x^i) \, , \, \ts_{ij}(\t, x^i) \, \}.
\ee
From this expression we read off the relation between the spacetime metric $\g_{\a\b}$ and the ADM fields 
\be\label{eq:ADMsplit}
\g_{\a\b}=\begin{pmatrix}
\eps\tN^2+\tN_i\tN^i & \tN_j\\
\tN_i & \ts_{ij}
\end{pmatrix} \, , \qquad
\g^{\a\b}=\begin{pmatrix}
\tfrac{1}{\eps \,\tN^2} & - \tfrac{\tN^j}{\eps \,\tN^2}\\
-\tfrac{\tN^i}{\eps \,\tN^2} \; \;  & \; \; \ts^{ij}+ \tfrac{\tN^i\tN^j}{\eps \,\tN^2} 
\end{pmatrix} \, .
\ee
Here the scalar products are with respect to the spatial metric $\ts_{ij}$. For completeness, we note that the decomposition of the determinant, appearing in the spacetime volume, is given as $\sqrt{\gamma} = \sqrt{\epsilon} \, \tN \, \sqrt{\ts}$.

\section{The RG equation for foliated spacetimes}
\label{sect.3}
In the sequel, we will adopt the viewpoint that the gravitational degrees of freedom are carried by the Lapse function $\tN$, the shift vector $\tN_i$ and the spatial metric $\ts_{ij}$ and derive the Wetterich equation for the component fields. We start with a discussion of symmetries and possible gauge fixings in subsection \ref{sect:3.1} before deriving the actual flow equation in subsection \ref{sect:3.2}. This construction will lead to a FRGE which intrinsically implements the foliation structure of spacetime and captures the RG flow of the projectable case of Ho\v{r}ava-Lifshitz gravity \cite{Horava:2008ih}.\footnote{As discussed in the introduction, the theory space underlying the non-projectable version of Ho\v{r}ava-Lifshitz gravity admits an additional class of interaction invariants constructed from the new building block $a_i = \p_i \tN/\tN$. Since in this case there are additional subtleties concerning the closure of the constraint algebra \cite{Henneaux:2009zb,Farkas:2010dw} it will not be discussed further at this point.} 
%
\subsection{Symmetries and gauge fixing}
\label{sect:3.1}
Since symmetries play a crucial role when constructing a FRGE, we start with a systematic discussion of the diffeomorphism symmetry in the ADM framework.
\subsubsection{Classical gravity in the ADM formalism}
Under a general coordinate transformation Diff($\cM$) the spacetime metric transforms according to $\gamma_{\a\b} \rightarrow \g_{\a\b} + \delta \g_{\a\b}$ with
\be\label{Difftrafo}
\delta \g_{\m\n} = \cL_v \g_{\m\n} \, .
\ee
Here $\cL_v$ denotes the Lie derivative of $\g$ with respect to the $D$-dimensional vector $v^\a(x^\m)$. The 
decomposition \eqref{decomp} then allows to write $v^\a$ in terms of its time and spatial parts $f(\t, x) \equiv t^\a v_\a$ and $\z^i(\t, x) = e_\a^i v^\a$ 
\be\label{vecdecADM}
v^\a = ( \, f(\t, x) \, , \, \z^i(\t, x) \, ) \, .
\ee 
Combining this split and the ADM decomposition \eqref{eq:ADMsplit} allows us to determine the transformation behavior of the component fields under Diff($\cM$)
\begin{align}\label{eq:gaugeVariations}
\d\tN &= \p_\t (f \tN ) + \z^k \p_k \tN  - \tN\tN^i\p_i f \, ,\nn\\
\delta \tilde N_i &= \partial_\tau(\tilde N_i f) + \zeta^k\partial_k\tilde N_i + \tilde N_k\partial_i\zeta^k
+ \tilde\sigma_{ki}\partial_\tau \zeta^k
+ \tilde N_k\tilde N^k\partial_i f  +\epsilon \tilde N^2\partial_i f \, , \\
\d\ts_{ij} &= f\p_\t \ts_{ij} + \z^k\p_k \ts_{ij} + \ts_{jk}\p_i\z^k + \ts_{ik}\p_j\z^k + \tN_j\p_i f + \tN_i\p_j f  \, . \nn
\end{align}
For completeness, we note 
\be\label{Nui}
\d\tN^i = \p_\t(\tN^i f) + \z^j\p_j \tN^i - \tN^j\p_j \z^i + \p_\t \z^i - \tN^i\tN^j\p_j f  + \eps \tN^2\ts^{ij}\p_j f \, .
\ee
One observes that, while Diff($\cM$) acts linearly on the metric $\gamma_{\m\n}$, the non-linearity of the ADM decomposition \eqref{eq:ADMsplit} leads to a non-linear transformation law for the component fields.

The gauge freedom can be exploited to adopt the proper-time gauge \cite{Dasgupta:2001ue}
\be\label{propertimegauge}
\tN(\t, x) = 1 \; , \qquad \tN^i(\t, x) = 0 \, .
\ee
This gauge choice still leaves the freedom to choose coordinate transformations which satisfy $\p_\t f = 0$ and $\p_\t\z^i + \eps \ts^{ik}\p_k f = 0$. Note that these equations encode the freedom to choose a coordinate system on the initial slice of the foliation. At the level of the path integral they are typically fixed by the corresponding boundary conditions, see \cite{Teitelboim:1981ua,Teitelboim:1983fk} for a more detailed discussion. Since we are not interested in surface effects, we will neglect this point in the subsequent discussion.

\subsubsection{Ho\v{r}ava-Lifshitz gravity: the projectable case}
Ho\v{r}ava-Lifshitz gravity \cite{Horava:2008ih} encodes the gravitational degrees of freedom in terms of the ADM fields. Considering the projectable version of the theory, the key difference is that here only the metric on the spatial slices and the shift vector are spacetime dependent fields, while the lapse function $\tN(\t)$ depends on time only and is constant along $\Sigma$
\be\label{gravmult}
\tilde{\chi}^{\rm pHL} \equiv \{ \, \tN(\t) \, , \, \tN_i(\t, x) \, , \, \ts_{ij}(\t, x) \, \}.
\ee
Moreover, the symmetry group is restricted to foliation-preserving diffeomorphisms Diff($\cM, \Sigma$). In this case the vector appearing in the Lie derivative \eqref{Difftrafo} is restricted to the form
\be\label{Folpresdiff}
v^\a(\t, x) = \left( \, f(\t) \, , \, \z^i(\t , x) \, \right) \, .
\ee

The variations of the component fields under foliation-preserving diffeomorphisms are those of \eqref{eq:gaugeVariations} restricted by the fact that $f$ and $\tN$ are space independent
\be\label{gaugeVariations2}
\begin{split}
\d\tN &= \p_\t(f \tN) \, , \\
\d\tN_i &= \p_\t (f\tN_i) + \z^k\p_k \tN_i + \tN_k\p_i \z^k + \ts_{ik} \p_\t \z^k    \, , \\
\d\ts_{ij} &= f\p_\t \ts_{ij} + \z^k\p_k \ts_{ij} + \ts_{jk}\p_i\z^k + \ts_{ik}\p_j\z^k \, , 
\end{split}
\ee
while
\be
\d\tN^i = \p_\t (f\tN^i) + \p_\t \z^i - \tN^k\p_k \z^i  + \z^k\p_k \tN^i \, .
\ee
In contrast to the full diffeomorphisms \eqref{eq:gaugeVariations} the foliation-preserving diffeomorphisms 
act linearly on the component fields. This will lead to considerable simplifications when applying the background-field method later on.

Again, we can use the gauge freedom to adopt the proper-time gauge \eqref{propertimegauge}. This fixes the Diff($\cM, \Sigma$) up to the residual transformations $\p_\t f = 0$ and $\p_\t \z^k = 0$, i.e., the choice of coordinate system on the initial slice. Thus, upon gauge-fixing, the field content of projectable Ho\v{r}ava-Lifshitz gravity and foliated QEG are identical. In this sense the transition from the field content \eqref{folQEG} and gauge-symmetries \eqref{vecdecADM} of the classical ADM formalism to \eqref{gravmult} and \eqref{Folpresdiff} can be understood as a partial gauge-fixing of the former. Owed to their identical off-shell field content, the Wetterich equation for the ADM formulation of classical gravity $\cT^{\rm fQEG}$ and projectable Ho\v{r}ava-Lifshitz gravity $\cT^{\rm pHL}$ will look identical. We stress, however, that the two theories are different in the sense that the latter allows a larger class of admissible interaction functionals, since the requirement of invariance under foliation-preserving diffeomorphisms is less restrictive then demanding invariance under Diff($\cM$). At this stage we proceed by constructing the Wetterich equation on $\cT^{\rm pHL}$, which retains Diff($\cM, \Sigma$) as a background symmetry. 

\subsubsection{Gauge fixing in the background field formalism}
The consistent quantization of the theory requires gauge fixing the symmetries \eqref{gaugeVariations2} in order to restrict the integration in the path integral to physically inequivalent configurations. For our purpose it is most convenient to implement this gauge fixing via the background field method \cite{back}, 
see \cite{Niedermaier:2006wt} for a detailed review. This construction ensures that the resulting
effective action contains interaction monomials that are invariant under \eqref{gaugeVariations2} only.

When implementing the background field method the quantum fields \eqref{gravmult} are split into fixed but arbitrary background fields $\bar{\chi} = \{ \bar N, \bar N_i, \bar{\sigma}_{ij} \}$ and fluctuations around 
this background $\hat{\chi} = \{ \hat N, \hat N_i, \hat{\sigma}_{ij} \}$
\be\label{bfsplitt}
\tN = \bar{N} + \hat{N} \, , \quad
\tN_i = \bar{N}_i + \hat{N}_i \, , \quad
\ts_{ij} = \bar{\sigma}_{ij} + \hat{\sigma}_{ij} \, . 
\ee
Note that the fluctuation fields are {\it not} assumed to be small in any sense. No expansion in powers of $\hat{\chi}$ is implied in this split.

The central element of the background field method is that the symmetry transformations
$\tilde{\chi} \mapsto \tilde{\chi} + \delta \tilde{\chi}$ can be distributed between the background
and fluctuation fields in different ways. Quantum gauge transformations leave the background invariant and attribute \eqref{gaugeVariations2} completely to the fluctuation field,
\be\label{qgt}
\bar{\chi} \mapsto \bar{\chi} \, , \qquad \hat{\chi} \mapsto \hat{\chi} + \delta_{\rm Q}(\bar{\chi} + \hat{\chi}) \, .
\ee
Here the variations $\delta_{\rm Q}$ are given by \eqref{gaugeVariations2} with $\tilde \chi$ substituted by
$\bar \chi + \hat \chi$. It is this set of symmetries that have to be gauge-fixed.

Background gauge transformations on the other hand act on both the background and fluctuation fields
\be\label{bgt}
\bar{\chi} \mapsto \bar{\chi} + \delta_{\rm B}\bar{\chi}  \, , \qquad \hat{\chi} \mapsto \hat{\chi} + \delta_{\rm B}  \hat{\chi} \, . 
\ee
This transformation plays the role of an auxiliary symmetry, which is retained by all the terms entering into the path integral. Its purpose is to ensure that the effective action contains interaction monomials that are invariant under Diff($\cM, \Sigma$) only. It thereby suffices that the background transformations become identical to the ones of the quantum field once the fluctuations are set to zero. This leaves some
freedom in the choice of $\delta_{\rm B}$. Explicitly, we adopt
\be\label{bgt2}
\begin{split}
\delta_{\rm B} \bar N =  \, \p_\tau \left( f \bar N \right) \, , \quad 
\delta_{\rm B} \bar N_i =  \, \p_\t \left( f \bar N_i \right) + \ts_{ij} \p_\t \zeta^j + \cL_\zeta(\bar N_i) \, , \quad 
\delta_{\rm B} \bar \s_{ij} = & f \p_\t \bar \s_{ij} + \cL_\zeta(\bar \s_{ij}) \,  
\end{split}
\ee
for the background fields while the fluctuations are taken to transform as
\be
\begin{split}
\delta_{\rm B} \hat N =  \, \p_\tau \left( f \hat N \right) \, , \quad
\delta_{\rm B} \hat N_i =  \, \p_\t \left( f \hat N_i \right)  + \cL_\zeta(\hat N_i) \, , \quad
\delta_{\rm B} \hat \s_{ij} = & f \p_\t \hat \s_{ij} + \cL_\zeta(\hat \s_{ij}) \, . 
\end{split}
\ee
Here $\mathcal{L}_\zeta$ denotes the Lie derivative on the spatial slices which  contains only spatial derivatives. Note that here the $\p_\tau \zeta$-terms have all been incorporated in the transformation 
of the background field. While this implies that the background fields in \eqref{bgt2} do not transform as the corresponding tensors, this still constitutes a choice of the background split since it reduces to the transformations \eqref{gaugeVariations2} in the limit of vanishing fluctuation fields. Combining these transformation laws shows that $\delta_{\rm B}\tilde \chi$ is given by \eqref{gaugeVariations2}. Here it is essential that Diff($\cM, \Sigma$) acts linearly on $\tilde{\chi}$. 

In the next step, we construct a gauge fixing term that implements the proper-time gauge in the background field formalism. By definition, this term has to break the quantum gauge transformation \eqref{qgt} while
being invariant under the background gauge transformation \eqref{bgt}. A straightforward computation then establishes that
\be\label{Sgf}
S^{\rm gf} = \frac{1}{2} \sqrt{\epsilon} \int d\t d^dx \, \bar{N} \sqrt{\bar{\sigma}} \, \left[ \frac{\hat{N}^2}{\alpha_L \, \bar{N}^2} + \frac{1}{\alpha_S \, \bar{N}^2} \, \hat{N}^i \, \bar{\sigma}_{ij} \, \hat{N}^j \right]
\ee 
indeed satisfied these requirements. In the Landau-limit, $\alpha_L \rightarrow 0, \alpha_S \rightarrow 0$, 
$S^{\rm gf}$ becomes a delta-distribution which eliminates the fluctuations of the lapse and shift vector, implementing the background proper-time gauge
\be\label{bptg}
F_{\rm lapse} = \frac{\hat{N}}{\bar N} = 0 \, , \qquad F_{\rm shift}^i \equiv \frac{\hat{N}^i}{\bar N} = 0 \, . 
\ee

The ghost action, exponentiating the Faddeev-Popov determinant arising from \eqref{Sgf} can then be found in
the standard way
\be\label{Sghost}
\begin{split}
S^{\rm gh} = \sqrt{\epsilon} \int d\t d^dx \,  \sqrt{\bar{\sigma}}
\bigg[ 
\bar{\omega} \p_\tau \tfrac{\tN}{\bar{N}} \omega
+ \bar{\omega}_i \left( \delta^i_j \p_\tau - \delta^i_j \tfrac{\tN^k}{\bN} \p_k + (\p_j \tfrac{\tN^i}{\bN}) \right) \omega^j 
\bigg] \, .
\end{split} 
\ee
The vector ghosts $\bar{\omega}_i(\tau, x), \omega^i(\tau, x)$ are functions of the spacetime, while the scalar ghosts $\bar{\omega}(\tau), \omega(\tau)$ depend on time only. The background lapse function $\bar{N}(\tau)$ has been distributed in such a way that the ghost action is invariant under background gauge transformations with the ghosts  transforming as scalars and vectors, respectively,
\be\label{backghosts}
\begin{split}
\delta_{\rm B} \omega = & f \, \p_\tau \omega \, , \qquad
\delta_{\rm B} \bar \omega_i =  f \, \p_\tau \bar \omega_i + \mathcal{L}_\zeta(\bar \omega_i) \, , \qquad
\delta_{\rm B} \omega^j =  f \, \p_\tau \omega^j + \mathcal{L}_\zeta( \omega^j) \, .
\end{split}
\ee
%

\subsection{The Functional RG Equation}
\label{sect:3.2}
After discussing the symmetries of our gravitational theory, we are now in the position to derive the Wetterich equation encoding the RG flow on $\cT^{\rm pHL}$. The construction follows the standard derivation \cite{Wetterich:1992yh,Reuter:1996cp}, see, e.g., \cite{Reuter:2007rv,Nagy:2012ef} for pedagogical reviews in the context of gravity.
\subsubsection{Defining the effective average action}
Our starting point is a generic action $S^{\rm grav}[\tN, \tN_i, \ts_{ij}]$ build from the multiplet \eqref{gravmult} and invariant under foliation-preserving diffeomorphisms \eqref{gaugeVariations2}. We then consider the {\it scale-dependent} generating functional $W_k$ for the connected Green functions
\be
\label{eq:partitionFunction}
\exp\{W_k[J; \bar{\chi}]\} = \int \cD\mu \, \exp\left[ - S^{\rm grav} - S^{\rm gf} - S^{\rm gh} - \Delta_kS - S^{\rm source} \right] \, .
\ee
Here $S^{\rm gf}$ and $S^{\rm gh}$ are given by \eqref{Sgf} and \eqref{Sghost}, respectively 
and the measure $\cD\mu$ consists of the integration over the gravitational fluctuations $\cD \hN \cD \hN_i \cD \hs_{ij}$ and the ghost contributions. The source term 
\be\label{Ssource}
\begin{split}
S^{\rm source} = 
& \, - \sqrt{\epsilon} \int d\t d^dx \bN \, \sqrt{\bar{\sigma}}  \left[ \frac{\tilde{t}}{\bar{N}^2} \hN  + \frac{\tilde{t}^i}{\bar{N}^2}  \hN_i  + t^{ij} \hs_{ij} +  \bar{\vartheta} \omega +  \vartheta \bar{\omega} + \bar{\vartheta}_i \omega^i + \vartheta^i \bar{\omega}_i \right] \, 
\end{split}
\ee
is invariant with respect to background-Diff($\cM, \Sigma$) if the sources transform covariantly with respect to time-reparametrizations $f$ and via the Lie derivative with respect to spatial diffeomorphisms $\zeta^i$. Here indices are raised and lowered with the spatial background metric. For convenience, we absorb the extra powers of $\bar{N}$ appearing in the lapse and shift terms into the sources, setting $t \equiv \tilde{t} \bar{N}^{-2}$ and $t^i \equiv \tilde{t^i} \bar{N}^{-2}$. The sources can then be collectively written as 
\be\label{Jsource}
J \equiv \{ \, t \, , \, t^i \, , \, t^{ij} \, , \, \bar{\vartheta} \, , \, \vartheta \, , \, \bar{\vartheta}_i \, , \, \vartheta^i \,  \} \, .
\ee

The essential piece in eq.\ \eqref{eq:partitionFunction} is the IR cutoff for the gravitational multiplet and the ghosts
\be\label{cutoff}
\begin{split}
\Delta_kS = & \, \frac{\sqrt{\epsilon}}{2} \int d\t d^dx \bar{N} \sqrt{\bs} \, \hat{\chi} \, \cR_k^{\rm grav}[\bar{\chi}] \, \hat{\chi} \\
& + \sqrt{\epsilon} \, \int d\t d^dx \, \bar{N} \sqrt{\bs} \, \left( \bar{\omega} , \bar{\omega}_i \right) \, \cR_k^{\rm gh}[\bar{\chi}] \, \left( \omega , \omega_j \right)^{\rm T} \, . 
\end{split}
\ee
The cutoff operators $\cR_k^{\rm grav}[\bar{\chi}]$ and $\cR_k^{\rm gh}[\bar{\chi}]$ serve the purpose of discriminating between the high-momentum and low-momentum fluctuations. Following \cite{Reuter:1996cp}, we use the eigenvalues of the Laplacian constructed from the {\it background} multiplet $\bar{\chi}$ to discriminate these modes. Eigenmodes of $-\bar{D}^2$ with eigenvalues $p^2 \gg k^2$ are integrated out without suppression whereas modes with small eigenvalues $p^2 \ll k^2$ are suppressed by a momentum dependent mass term. Generally, the $\cR_k$ have the structure
\be\label{profile}
\cR_k[\bar{\chi}] = \cZ_k R_k = \cZ_k \, k^2 \, R^{(0)}(-\bar{D}^2/k^2) \, . 
\ee
Here $\cZ_k$ is a background-field dependent matrix that ensures the invariance of \eqref{cutoff} with respect to the background-gauge transformations. The dimensionless shape function $R^{(0)}$ interpolates between $R^{(0)}(0) = 1$ and $\lim_{w \rightarrow \infty} R^{(0)}(w) = 0$. Convenient choices are, e.g., the exponential cutoff $R^{(0)}(w) = w [\exp(w) -1]^{-1}$ or the optimized cutoff \eqref{Ropt}.  

In the next step, we construct the $k$-dependent classical fields 
\be\label{classfld}
\chi = \{ \, h \, , \, h_i \, , \, h_{ij}  \, , \, \bar{c} \, , \, c \, , \, \bar{c}_i \, , \, c^i \} 
\ee
as the expectation value of the fluctuation fields. These are easily found as variations of $W_k$ with respect to the sources
\be\label{gravclass}
\chi^a = \frac{1}{\sqrt{\epsilon} \bN \sqrt{\bs}} \frac{\delta W_k}{\delta J_a} \, .
\ee
The extra factors $\bN$ included in \eqref{Jsource} ensure that the classical lapse and shift transform as the corresponding quantum fields
\be\label{classtrafo}
\begin{split}
\delta_{\rm B} h = &  \p_\tau (f \, h) \, , \qquad
\delta_{\rm B} h_i =  \p_\tau (f h_i) + \mathcal{L}_\zeta(h_i) \, . 
\end{split}
\ee

As usual, we assume that one can invert the relations \eqref{gravclass} and solve for the sources as functionals of the classical fields and, parametrically, of the background. The Legendre transform $\tilde \Gamma_k$ of $W_k$ reads
\be\label{tgamma}
\tilde \Gamma_k[\chi; \bar{\chi}] = \sqrt{\epsilon} \int d\t d^dx \bar{N} \sqrt{\bs} \left[ t h + t^i h_i + t^{ij} h_{ij} 
+ \bar{\vartheta} c +  \vartheta \bar{c} + \bar{\vartheta}_i c^i + \vartheta^i \bar{c}_i \right] - W_k \, . 
\ee
The effective average action $\Gamma_k$ is then obtained from $\tilde \Gamma_k$ by subtracting the cutoff action $\Delta_k S$ with the classical fields inserted
\be\label{EAAdef}
\Gamma_k[\chi; \bar{\chi}] = \tilde \Gamma_k[\chi; \bar{\chi}] - \Delta_kS[\chi; \bar{\chi}] \, . 
\ee

\subsubsection{The Wetterich equation}
The derivation of the Wetterich equation encoding the $k$-dependence of $\Gamma_k$ starts with the connected two-point function
\be\label{2ptfct}
(G_k)^{ab}(x,\t_x ;y,\t_y)=\frac{1}{\eps \bN^2 \sqrt{\bs(x,\t_x)\bs(y,\t_y)}}\frac{\d^2W_k}{\d J_a(x,\t_x)\d J_b(y,\t_y)} \, ,
\ee
which is matrix-valued in field space. Correspondingly, the Hessian $\tilde\G_k^{(2)}$ of \eqref{tgamma} is given by
\be\label{hessian}
\left( \tilde\G_k^{(2)} \right)_{ab} := (-1)^{[b]}\frac{1}{\eps \, \bar{N}^2 \, \sqrt{\bs(x,\t_x) \, \bs(y,\t_y)}}\frac{\d^2\tilde\G_k}{\d\chi^a(x,\t_x)\d\chi^b(y,\t_y)} \, ,
\ee
where the index $[b]$ takes the value zero for commuting and one for anti-commuting fields $\chi_b$, respectively. The Legendre transform \eqref{tgamma} implies that \eqref{2ptfct} and \eqref{hessian} are each others inverse in a functional sense
\be \label{eq:inverse}
\sqrt{\eps} \int d^dyd\t_y \, \bN \sqrt{\bs} \, (G_k)^{ac}(x,\t_x ;y,\t_y)\left( \tilde\G_k^{(2)} \right)_{cb}(y,\t_y ;z,\t_z) = \frac{\d^a_b\d(x-z)\d(\t_x - \t_z)}{\sqrt{\eps} \bN \sqrt{\bs}} \, .
\ee
Taking the derivative of \eqref{eq:partitionFunction} with respect to the RG ``time'' $t = \ln(k)$ then gives
\be\label{eq:WDerivative}
-\p_t W_k = \frac{1}{2} \, \mathrm{Tr} \, \Big[ \, \langle \hat{\chi}^a \otimes \hat{\chi}^b \rangle \,  \p_t \cR_k|_{ab} \, \Big] \, .
\ee
Here $\langle \hat{\chi}^a \otimes \hat{\chi}^b \rangle$ denotes the two-point correlator of the fluctuation fields (including the ghost fields) and $\mathrm{Tr}$ includes an integration over loop momenta and a sum over internal indices. The two-point correlator is related to the connected two-point function $(G_k)_{ab}$ and the classical fields, \eqref{gravclass}, via
\be
\begin{split}
\langle \hat{\chi}^a \otimes \hat{\chi}^b \rangle = & \, (G_k)^{ab} + \chi^a \, \chi^b \, 
=  \, \left( \tilde\G_k^{(2)} \right)_{ab}^{-1} + \chi^a \, \chi^b \, . 
\end{split} 
\ee
Expressing the flow \eqref{eq:WDerivative} in terms of the effective average action \eqref{EAAdef} then yields the desired functional renormalization group equation for Ho\v{r}ava-Lifshitz gravity
\be\label{eq:wetterichEquation}
\p_t \G_k[\chi; \bar{\chi}] = \frac{1}{2}\mathrm{STr}\left[ \left( \G_k^{(2)} + \cR_k \right)^{-1} \p_t \cR_k \right] \, .
\ee
Here the Hessian $\G_k^{(2)} $ denotes the second derivative of $\Gamma_k[\chi; \bar{\chi}]$ with respect to the fields $\chi$. Both $\G_k^{(2)} $ and $\cR_k$ are matrix valued in field space and the supertrace includes an integral over loop momenta and a sum over field space. The FRGE \eqref{eq:wetterichEquation} constitutes the central result of this section.

We close this subsection by discussing the limits of the RG flow implied by \eqref{eq:wetterichEquation}. Since the $\cR_k$'s vanish for $k=0$, the limit $k \rightarrow  0$ of $\Gamma_k[\chi; \bar{\chi}]$ brings us back to the standard effective-action functional which still depends on two sets of fields $\chi$ and $\bar{\chi}$. The ordinary effective action $\Gamma[\bar{\chi}]$ with one argument is obtained from this functional by setting the expectation value of the fluctuation fields $\chi = 0$ \cite{back}
\be
\Gamma[\bar{\chi}] = \lim_{k \rightarrow 0} \Gamma_k[\chi = 0; \bar{\chi}] \, . 
\ee

Besides the FRGE \eqref{eq:wetterichEquation} the effective average action also satisfies an exact integro-differential equation, which can be used to find the $k \rightarrow \infty$ limit of the average action \cite{Reuter:1996cp}:
\be
\Gamma_{k \rightarrow  \infty}[\chi ;\bar \chi] = 
S[\bar{\chi}+\chi] + S_{\rm gf}[\chi; \bar{\chi}] 
+S_{\rm gh}[\chi; \bar{\chi}]\,.
\label{F24}
\ee
Intuitively, this limit can be understood from the observation that for $k \rightarrow \infty$ all quantum fluctuation in the path integral are suppressed by an infinite mass term. Thus, in this limit no fluctuations are integrated out and $\Gamma_{k \rightarrow  \infty}$ agrees with the microscopic action $S$ supplemented by the gauge fixing and ghost actions, also see \cite{Manrique:2008zw} for more details.

\subsubsection{Symmetries preserved by the RG flow}
The main advantage in the use of the background field method is that the functional $\tilde \G_k$ and, as a result, also $\G_k$ is invariant under background foliation-preserving diffeomorphisms, when all its arguments transform according to their transformation rules
\be\label{inv}
\Gamma_k[\chi + \delta_B \chi; \bar{\chi} + \delta_B \bar{\chi}] = \Gamma_k[\chi; \bar{\chi}] \, . 
\ee
Note that here, contrary to the quantum-gauge transformation \eqref{qgt}, also the background fields transform according to their corresponding symmetries. The invariance \eqref{inv} is a consequence of
\be
W_k[J + \delta_B J; \bar{\chi} + \delta_B \bar{\chi}] = W_k[J; \bar{\chi}] 
\ee
which in turn follows from \eqref{eq:WDerivative} if one uses the invariance of $S^{\rm grav}$, $S^{\rm gf}$, $S^{\rm gh}$ and $\Delta_kS$ under the background transformations \eqref{bgt} and \eqref{backghosts}. At this point, we assume that the functional measure in \eqref{eq:WDerivative} is invariant under Diff($\cM, \Sigma)$.

The background-gauge invariance of $\Gamma_k$, expressed in eq.\ \eqref{inv}, is of enormous practical importance. It implies that a RG flow, starting from a background foliation-preserving diffeomorphism invariant $\Gamma_{\hat k}$ at a scale $\hat k$, will not generate interactions that violate this background symmetry dynamically. Nevertheless, even if the initial action is simple, the RG flow will generate all sorts of local and non-local terms in $\Gamma_k$ which are consistent with the symmetries. 

We close this section with the following important remark. Owed to the requirement that the cutoff $\Delta_kS$ must be quadratic in the fluctuation fields, it generically seems impossible to construct a Wetterich-type FRGE which employs a linear background split and preserves a non-linear symmetry as a background symmetry. In principle it should be possible, however, to construct a FRGE for the gravitational multiplet \eqref{folQEG} 
where the full Diff($\cM$) is realized as a background symmetry. Since the corresponding symmetry transformations \eqref{eq:gaugeVariations} are non-linear, we expect that this construction will either require a generalization of the background gauge fixing procedure to non-linear symmetries \cite{Kazakov:1977mw} or a non-linear background split \cite{Niedermaier:2006wt}. From the flow equation for gravity in the metric formalism \cite{Reuter:1996cp}, we expect that this will lead to a functional renormalization group equation on a subspace $\cT^{\rm fQEG} \subset \cT^{\rm pHL}$ which is {\it closed under RG transformations}. While constructing the corresponding flow equation is certainly interesting we will not embark on this construction at this point.

\section{Signature effects in the gravitational RG flow}
\label{sect:4}
In principle the Wetterich equation \eqref{eq:wetterichEquation}  constitutes an exact RG equation on $\cT^{\rm pHL}$ which is, however, notoriously difficult to solve. A common technique to find approximate solutions of the equation which do not rely on an expansion in a small parameter consists in truncating the effective average action by restricting $\G_k$ to a finite set of running coupling constants. In this section we will derive the beta functions in the simplest gravitational setting, the foliated Einstein-Hilbert truncation. This setup allows the direct comparison of the approximate RG flows on $\cT^{\rm mQEG}$ and $\cT^{\rm fQEG}$. Owed to the foliated background, our derivation uncovers many structures that are well-known  in the context of quantum field theory at finite temperature \cite{Litim:1998yn,Litim:2006ag,Floerchinger:2011sc}. 

\subsection{The ADM-decomposed Einstein-Hilbert truncation}
The ansatz for $\G_k$ corresponding to the ADM-decomposed Einstein-Hilbert truncation is of the general form\footnote{In the terminology of \cite{MRS} this corresponds to a single-metric computation. Results for metric flows that take into account the effect of higher order terms in the fluctuation fields have recently been reported in \cite{Donkin:2012ud,Christiansen:2012rx}.}
\be\label{ansatz}
\G_k[\chi; \bar{\chi}] \approx \G_k^{\rm grav}[\chi + \bar{\chi}] + S^{\rm gf}[\chi; \bar{\chi}] + S^{\rm gh}[\chi; \bar{\chi}]
\ee
where we have approximated the gauge-fixing and ghost part of the effective average action by their classical expressions \eqref{Sgf} and \eqref{Sghost}. For the gravitational part $\G_k^{\rm grav}$ we adopt the Einstein-Hilbert action in $D = d+1$ spacetime dimensions. Expressing
\be
\G_k^\mathrm{grav} = \frac{1}{16\pi G_k} \int d^Dx \sqrt{g} \left[ -R +  2\Lambda_k \right]
\ee
in terms of the gravitational multiplet \eqref{gravmult} the corresponding action reads
\be \label{eq:decomposedEHaction}
\G_k^\mathrm{grav} = \frac{\sqrt{\eps}}{16\pi G_k} \int d\t d^{d}x \,N\sqrt{\s}\left[ \eps^{-1} \,    \left( K_{ij} K^{ij} - K^2 \right)  - {}^{(d)}R + 2\Lambda_k \right] \, .
\ee
Here the $\{ N, N_i, \s_{ij} \}$ are the classical counterparts of the full quantum fields $\tilde \chi$
\be\label{classmult}
N \equiv \bar{N} + h \, , \qquad N_i \equiv \bar{N}_i + h_i \, , \qquad \sigma_{ij} \equiv \bs_{ij} + h_{ij} \, .   
\ee
Moreover
\be
 K_{ij}=\frac{1}{2N}\left( \p_\t{\s}_{ij}- D_i N_j-D_j N_i \right)
\ee
is the extrinsic curvature, $K \equiv \sigma^{ij} K_{ij}$, and $D_i$ and ${}^{(d)}R$ denote the covariant derivative and the intrinsic curvature of the $d$-dimensional spatial slices constructed from $\s_{ij}$, respectively. The functional form \eqref{eq:decomposedEHaction} can easily be established by substituting the ADM decomposition \eqref{eq:ADMsplit} into the standard Einstein-Hilbert action written in terms of the covariant metric $\gamma_{\a\b}$. As in the standard Einstein-Hilbert case \eqref{eq:decomposedEHaction} contains two scale-dependent coupling constants, the Newton constant $G_k$ and the cosmological constant $\Lambda_k$.

The action \eqref{eq:decomposedEHaction} is, by construction, invariant under the full diffeomorphism group Diff($\cM$). This symmetry fixes the relative coefficients between the extrinsic and intrinsic curvature terms to the form \eqref{eq:decomposedEHaction}. When restricting the symmetry group to the foliation-preserving diffeomorphisms \eqref{gaugeVariations2}, the $K^2$ term, the $K^{ij}K_{ij}$ term and the $R$ term are invariant on their own. Thus the anisotropic case of Ho\v{r}ava-Lifshitz gravity allows to introduce further coupling constants in front of the terms. The study of their RG flow will be subject to a future publication \cite{inprep}.

\subsection{Constructing the functional traces}
Substituting the ansatz \eqref{ansatz} into the FRGE \eqref{eq:wetterichEquation}, we observe from its left hand side that the scale dependence of $G_k$ and $\Lambda_k$ can be read off from the coefficients of the spacetime volume and the curvature on the spatial slices ${}^{(d)}R$. Thus it is sufficient to project the traces appearing on the right hand side onto these two curvature invariants. We stress that when computing the beta functions of the theory the geometric quantities merely act as a bookkeeping devices. No physical meaning should be attached to them.

The first step in evaluating the functional traces arising from \eqref{ansatz} constitutes in computing the Hessian $\G_k^{(2)}[\chi, \bar \chi]$ with respect
to the fluctuation fields. This calculation can be simplified in a number of ways. Firstly, we adopt Landau gauge by sending $\alpha_L \rightarrow 0$, $\alpha_S \rightarrow 0$. In this limit the gauge-fixing term \eqref{Sgf} is converted to a $\delta$-function which freezes the fluctuation fields $\hat N = 0$, $\hat N_i = 0$. Thus $\hat N$ and $\hat N_i$ decouple and do not contribute to the traces on the right hand side of the flow equation. Thus,
\be
\begin{split}
\p_t \G_k = \half {\rm Tr} \Bigg[ \left( \frac{\delta^2 \G_k^{\rm grav}}{\delta h_{ij} \delta h_{kl}} + \cR_k^{\rm grav} \right)^{-1} \p_t \cR_k^{\rm grav} \Bigg]
- {\rm Tr} \Bigg[ \left( \cM + \cR_k^{\rm gh} \right)^{-1} \p_t \cR_k^{\rm gh} \Bigg]
\end{split}
\ee
where the Hessian $\cM$ in the ghost trace is given by
\be
\cM = 
{\rm diag}\left[
\frac{\delta^2 S^{\rm gh}}{\delta c \delta \bar{c}} \, , \, 
\frac{\delta^2 S^{\rm gh}}{\delta c^i \delta \bar{c}^j}  \, 
\right] \, .
\ee

The second simplification originates from the observation that we can set the fluctuation fields $\chi=0$ once the second variations are computed. Thus we can use the background covariance of the construction to carry out the computation in a specific background geometry. The choice of background has to be general enough to distinguish the two interaction monomials carrying the information of the RG flow. A natural choice uses
\be\label{background}
\overline{\cM} = S^1 \times S^d \, ,
\ee
i.e., the direct product of a ``time''-circle with periodicity $T$ and a time-independent maximally symmetric sphere $S^d$. In terms of the background multiplet, this choice implies
\be\label{back2}
\bar{N}(\t) = 1 \, , \qquad \bar N_i(\t, x) = 0 \, , \qquad \bs_{ij}(\t, x) = \bs_{ij}(x)|_{S^d} \, .
\ee
Owed to the maximal symmetry of $S^d$ the background curvatures on the spatial slices satisfy
\be\label{eq:backsphere}
\partial_\t \bs_{ij} = 0 \, , \qquad \Rb_{ij} = \frac{1}{d} \, \Rb \, \bs_{ij} \, , \qquad \Rb_{ijkl} =  \frac{\Rb}{d(d-1)} \, \left( \bs_{ik} \, \bs_{jl} - \bs_{il} \, \bs_{jk} \right) \, . 
\ee
In order to lighten our notation, we will drop the prefix $(d)$ of the intrinsic curvature from now on.

We then proceed to compute the Hessian of $\Gamma_k^{\rm grav}$ around the background \eqref{background}. Inspecting \eqref{eq:decomposedEHaction}, there are four distinguished interaction monomials entering into $\Gamma_k^{\rm grav}$
\be
\begin{array}{ll}
I_1 = \int d\tau d^dx \, N \,  \sqrt{\sigma} \, , \qquad & I_2 = \int d\tau  d^dx \, N \,  \sqrt{\sigma} \;  R \, , \\[1.2ex]
I_3 = \int d\tau d^dx \, N \, \sqrt{\sigma}  \, K_{ij} K^{ij}  \, , \qquad & I_4 = \int d\tau  d^dx \, N \, \sqrt{\sigma} \, K^2  \, .
\end{array}
\ee
Using the background \eqref{back2}, \eqref{eq:backsphere} to simplify the result, the second variations of these monomials read
\be\label{var1}
\begin{split}
\delta^2 I_1 = & \int d\tau  d^dx \sqrt{\bs} \left\{ - \half \, h^{\mathrm{T}ij} \, h^\mathrm{T}_{ij} + \tfrac{d-2}{4d} \, h^2 \right\} \, ,\\
\delta^2 I_2 = & \int d\tau  d^dx \sqrt{\bs} \bigg\{
h \left[ \tfrac{(d-2)(d-1)}{2d^2} \Delta + \tfrac{(d-2)(d-4)}{4d^2} \Rb \right] h - \half h^\mathrm{T}_{ij} \left[ \Delta + \tfrac{d^2-3d+4}{d(d-1)} \Rb \right] h^{\mathrm{T}ij} \\ & \qquad \qquad \qquad \quad
+ \tfrac{d-2}{d} h \Db_i \Db_j h^{\mathrm{T}ij} - h^{\mathrm{T}ik} \Db_k \Db_l h^{\mathrm{T}lj}\bar\sigma_{ij}
\bigg\} \, , \\
\delta^2 I_3 = & - \tfrac{1}{2} \, \int d\tau  d^dx \sqrt{\bs} \, \Big\{  h^\mathrm{T}_{ij} \, \p_\tau^2 \,  h^{\mathrm{T}ij} + \tfrac{1}{d} \, h  \, \p_\tau^2 \, h \Big\} \, , \\
\delta^2 I_4 = & - \tfrac{1}{2} \, \int d\tau  d^dx \sqrt{\bs} \,  h \, \p_\tau^2 \, h  \, .
\end{split}
\ee
Here we decomposed the fluctuations $h_{ij}$ into their traceless and trace part,
\be
h_{ij} = h_{ij}^ {\rm T} + \frac{1}{d} \bs_{ij} \, h \, , \, \qquad \bs^{ij} h_{ij}^{\rm T} = 0 \,  
\ee
and $\Delta = - \bs^{ij} \bar{D}_i \bar{D}_j$ is the background Laplacian built from the spatial metric. Here and in the following the $h$ denotes the trace-part of the classical fluctuation field $ h_{ij}$ and should not be confused with the classical expectation value of the lapse function, \eqref{gravclass}, which has been eliminated by the proper-time gauge choice \eqref{bptg}.

A technical difficulty arises from the fact that in \eqref{var1} not all covariant derivatives combine into Laplace operators. In particular the second line in $\delta^2 I_2$ leads to a so-called non-minimal operator structure in $\Gamma^{(2)}_k$.\footnote{This feature was already expected based on earlier computations of the beta functions in the metric formulation \cite{Reuter:1996cp,Lauscher:2001ya,Reuter:2001ag}, where a special gauge choice is required in order to remove these non-minimal terms.} Following \cite{Lauscher:2001ya,Benedetti:2010nr}, we bypass this problem by further decomposing the traceless part of the metric fluctuations $h^{\rm T}_{ij}$ into its irreducible representations on the sphere via the transverse-traceless (TT) decomposition \cite{York}
\be\label{TTdec}
h_{ij}^{\rm T} = h_{ij}^{\rm TT} + 2 \Db_{(i}  \big[ \Delta - \tfrac{1}{d} \Rb  \big]^{-1/2} \xi_{j)}
 + \big[ \Db_i \Db_j - \tfrac{1}{d} \bs_{ij} \big] \big[ \Delta (\Delta - \tfrac{1}{d-1} \Rb ) \big]^{-1/2} \varsigma \, . 
\ee
Here the component fields are subject to the differential constraints
\be\label{divconst}
\bs^{ij} h_{ij}^{\rm TT} = 0 \, , \qquad \Db^i h_{ij}^{\rm TT} = 0 \, , \qquad \Db^i \xi_i= 0 \, .
\ee
Moreover, we have chosen the normalization of the transverse vector and the scalar $\varsigma$ such that the TT decomposition \eqref{TTdec} substituted into the Gaussian integral $\int \cD h_{ab} \exp[- \int d^dx \sqrt{\bs} h_{ij} h^{ij}]$ does not give rise to operator-valued determinants. Thus the normalized TT decomposition \eqref{TTdec} does not require the introduction of auxiliary fields exponentiating the Jacobi determinants naturally appearing in this type of field redefinitions as, e.g., implemented in \cite{Machado:2007ea}. 

Substituting the TT decomposition \eqref{TTdec} yields the second variation in terms of the component fields
\be\label{variations}
\begin{split}
\delta^2 I_1 = & \int d\tau d^dx \sqrt{\bs} \left\{ - \half h^{ij\mathrm{TT}} \, h^\mathrm{TT}_{ij} - \tfrac{d-1}{2d} \varsigma^2 - \xi^j \, \xi_j + \tfrac{d-2}{4d} \, h^2 \right\} \, ,\\
\delta^2 I_2 = & \int d\tau d^dx \sqrt{\bs} \bigg\{ \half C_0 h \left[ \Delta + \tfrac{d-4}{2(d-1)} \Rb \right] h 
- \half h^\mathrm{TT}_{ij} \left[ \Delta + C_{\rm 2T} \Rb \right] h^{ij\mathrm{TT}} 
+ \half C_0 \varsigma\left[ \Delta - \Rb \right]\varsigma  \\ & \qquad \qquad \qquad \quad
- \tfrac{d-2}{d}\Rb \, \xi_i \,\xi^i + C_0 \, h \left[ \Delta^2 - \tfrac{1}{d-1} \Rb \Delta \right]^{1/2} \varsigma
\bigg\} \, , \\
\delta^2 I_3 = & - \tfrac{1}{2} \, \int d\tau d^dx \sqrt{\bs} \, \Big\{  h^{ij\mathrm{TT}} \p_\tau^2 h^\mathrm{TT}_{ij} + \tfrac{d-1}{d}\varsigma \, \p_\tau^2 \, \varsigma + 2\xi^j \, \p_\tau^2 \, \xi_j + \tfrac{1}{d} \, h  \, \p_\tau^2 \, h \Big\} \, , \\
\delta^2 I_4 = & - \tfrac{1}{2} \, \int d\tau d^dx \sqrt{\bs} \,  h \, \p_\tau^2 \, h  \, .
\end{split}
\ee
Here the coefficients $C_i$ are functions of the dimension of the spatial slices $d$
\be\label{eq:abbreviations}
C_{\rm 2T} \equiv \frac{d^2-3d+4}{d(d-1)} \, , \qquad C_0 \equiv \frac{(d-2)(d-1)}{d^2} \, .
\ee
In order to derive \eqref{variations} from \eqref{var1} it is useful to first carry out the TT-decomposition without including the normalization factors for the transverse vector and scalar $\varsigma$ and subsequently noticing that the normalization factors appear as factors in the hessians. Moreover, we used that the background metric is $\tau$-independent, such that all background quantities (including the spatial Laplacians) commute with the $\t$ derivatives, which leads to further drastic simplifications.

Based on \eqref{variations} it is now straightforward to write the part of $\G_k^{\rm grav}$ quadratic in the fluctuation fields
\be
\Gamma_k^{\rm grav}[\sigma_{ij}] =  \Gamma_k^{\rm grav}[\bs_{ij}] + \cO(h) + \half \, \delta^2 \Gamma_k^{\rm grav}[h_{ij}; \bs_{ij}] \, . 
\ee
It is convenient to list the contributions appearing in $ \delta^2 \Gamma_k^{\rm grav}[h_{ij}; \bs_{ij}]$ according to the components of the fluctuation fields contained
\be\label{g2varr}
\begin{split}
\delta^2 \G_{\rm TT}^{\rm grav} = & \, \frac{\sqrt{\eps}}{16 \pi G_k} \int d\t d^dx \sqrt{\bs} \, \half h_{ij}^{\rm TT} \, \left[ \Delta - \tfrac{1}{\eps} \p_\t^2 + C_{\rm 2T} \Rb - 2 \Lambda_k \right] \, h^{{\rm TT}ij} \, , \\
\delta^2 \G_{ \xi\xi}^{\rm grav} = & \, \frac{\sqrt{\eps}}{16 \pi G_k} \int d\t d^dx \sqrt{\bs} \, \xi_i \left[ - \tfrac{1}{\eps} \p_\t^2 + \tfrac{d-2}{d} \Rb - 2 \Lambda_k \right] \xi^i \, , \\
\delta^2 \G_{\varsigma\varsigma}^{\rm grav} = & \, \frac{\sqrt{\eps}}{16 \pi G_k} \int d\t d^dx \sqrt{\bs} \, \half \varsigma \left[
- C_0 (\Delta - \Rb ) - \tfrac{d-1}{d\eps} \p_\t^2 - \tfrac{2(d-1)}{d} \Lambda_k 
\right] \varsigma \, , \\
\delta^2 \G_{hh}^{\rm grav} = & \, \frac{\sqrt{\eps}}{16 \pi G_k} \int d\t d^dx \sqrt{\bs} \, \half h \left[
- C_0 \Delta + \tfrac{d-1}{d\eps} \p_\t^2 - \tfrac{(d-2)(d-4)}{2d^2} \Rb + \tfrac{d-2}{d} \Lambda_k \right] h \, , \\
\delta^2 \G_{h\varsigma}^{\rm grav} = & \, - \frac{\sqrt{\eps}}{16 \pi G_k} \int d\t d^dx \sqrt{\bs} \, C_0 h \, \left[ \Delta^2 - \tfrac{1}{d-1} \Rb \Delta
\right]^{1/2} \, \varsigma \, .
\end{split}
\ee

These variations are complemented by the analogous variations appearing in the ghost sector. Setting the background ghost fields to zero, which suffices for the present computation, \eqref{Sghost} leads to second variations that are diagonal in the ghost fluctuations. Explicitly, they read
\be\label{ghostvar}
\begin{split}
\delta^2 \G_{\bar{c}c}^{\rm grav} = & \, \sqrt{\eps} \int d\t d^dx \sqrt{\bs} \, \bar{c} \p_\t c \, , \\
\delta^2 \G_{\bar{c}^ic^j}^{\rm grav} = & \, \sqrt{\eps} \int d\t d^dx \sqrt{\bs} \, \bar{c}^i \p_\t c_i \, . \\
\end{split}
\ee

Based on the variations \eqref{g2varr} and \eqref{ghostvar} we are now in a position to specify the IR regulators $\cR_k$. As a key requirement $\cR_k$ must be positive definite for all values of $k$, in order to act as a $k$-dependent mass term. In general, this will not be the case if $\Delta$ is a Laplacian constructed from a Lorentz signature metric. Since one of  our interests is in the signature effects of the spacetime on the gravitational RG flow we choose a purely spatial regulator, i.e., $\cR_k = \cR_k(\Delta)$ will be a function of the Laplacian on the spatial slice only. In this case $\Delta$ is positive semi-definite. Moreover, this choice is sufficient to regularize the operator traces appearing in the flow equation \cite{Litim:2006ag}.\footnote{Covariant cutoffs have recently been proposed in \cite{Floerchinger:2011sc}. Since their implementation in the Lorentzian setting is slightly involved we will resort to the spatial regulators at this stage. It will be interesting to carry out the expansion around the mass poles of the Lorentzian propagators and compare to the results obtained here.} 

In practice we will implement a regulator of Type I, in the nomenclature of \cite{Codello:2008vh}, dressing the spatial Laplacians with a $k$-dependent mass term according to
\be\label{TypeIrule}
\Delta \mapsto \Delta + k^2 \, R_k^{(0)}(\Delta/k^2) \equiv P_k \, .
\ee
Here $k^2$ are purely spatial momenta and $R^{(0)}$ is the profile function introduced in \eqref{profile}. The rule \eqref{TypeIrule} fixes the matrices $\cZ_k$ and we obtain
\be\label{cutoffs}
\begin{split}
\cR_{\rm TT} = & \frac{1}{32 \pi G_k} \mathbbm{1}_{\rm 2T} R_k \, , \\ 
\cR_{hh} = \cR_{\varsigma\varsigma} = & - \frac{1}{32 \pi G_k} C_0 R_k \, , \\
\cR_{\varsigma h} = \cR_{h \varsigma} = & - \frac{1}{32 \pi G_k} \, C_0 \, \left[ \left(P_k^2 - \tfrac{1}{d-1} \Rb P_k \right)^{1/2} - \left(\Delta^2 - \tfrac{1}{d-1} \Rb \Delta \right)^{1/2} \right]\, .
\end{split}
\ee
The quadratic variations of the transverse vector $\xi$ and the ghosts do not contain spatial Laplacians. Thus the corresponding regulators $\cR_k$ constructed via \eqref{TypeIrule} vanish. As a consequence, they do not source the RG flow of the Newton constant and the cosmological constant in the present truncation.

We now have all the ingredients to write down the explicit form of the traces contributing to the flow of the truncation ansatz \eqref{eq:decomposedEHaction}. Here it is convenient to separate the contributions of the transverse-traceless and scalar sectors according to
\be\label{fl2}
\p_t \Gamma_k = \cT_{\rm TT} + \cT_{\rm s,1} + \cT_{\rm s,2}  \, . 
\ee
Explicitly, these traces are given by
\be\label{tensortraces}
\cT_{\rm TT} = \half \, (32 \pi G_k) \,  {\rm Tr} \left[ \left( P_k - \tfrac{1}{\epsilon} \p_\t^2 + C_{\rm 2T} \Rb - 2 \Lambda_k \right)^{-1} \, \p_t \cR_{\rm TT} \right]
\ee
and
\be\label{scalartraces}
\begin{split}
\cT_{\rm s,1} = &  \half \, (32 \pi G_k) \,  {\rm Tr} \left[ \frac{1}{\cN}  \, \left( - 2 C_0 P_k + \tfrac{d^2-4}{2d^2} \Rb - \Lambda_k \right) \, \p_t \cR_{\rm hh} \right] \, ,  \\
\cT_{\rm s,2} = &  \half \, (32 \pi G_k) \,  {\rm Tr} \left[ \frac{1}{\cN} \, 2 C_0 \, \left( P_k^2 - \tfrac{1}{d-1} \Rb P_k  \right)^{1/2} \, \p_t \cR_{h\varsigma} \right] \, .
\end{split}
\ee
Here the cutoff functions are given in \eqref{cutoffs} and $\cN$ is the determinant of the two-by-two matrix appearing in the scalar sector
\be
\begin{split}
\cN = & C_0  \Lambda_k \left( P_k - 2  \Lambda_k \right) - \tfrac{(d-1)^2}{d^2} \p_\t^4 - \tfrac{d-1}{\epsilon d^2} (3d-4)  \Lambda_k \p_\t^2 \\
& - \tfrac{C_0}{2d} (d-2) \Rb \left( P_k - \tfrac{3}{\epsilon} \p_\t^2 + \tfrac{d-4}{d} \Rb - \tfrac{4(d-3)}{d-2}  \Lambda_k\right) \, . 
\end{split}
\ee
The traces $\cT_{\rm s,1}$ and $\cT_{\rm s,2}$ thereby capture the contribution from the diagonal and off-diagonal part of the scalar field matrix, respectively.

For the computation of the RG flow, it is sufficient to keep track of the intrinsic curvature scalar up to the first order in $\Rb$, since all higher powers of $\Rb$ are outside the truncation subspace. This feature can be used to combine the two scalar traces by expanding the numerator in $\cT_{\rm s,2}$ around $\Rb = 0$. As a result, the two scalar contributions combine into
\be\label{scalartrace2}
\cT_{\rm s} \equiv   \cT_{\rm s,1} + \cT_{\rm s,2} \simeq \half \, (32 \pi G_k) \,  {\rm Tr} \left[ \frac{1}{\cN} \left( \tfrac{d-2}{2d} \Rb - \Lambda_k \right) \p_t \cR_{hh} \right] + \cO(\Rb^2)
\ee
which considerably simplifies the evaluation of this sub-sector in the next subsection.

\subsection{Constructing the beta functions}
The projection of the traces \eqref{tensortraces} and \eqref{scalartraces} onto the truncation subspace conveniently exploits that the background has the structure $S^1 \times S^d$. This structure allows to Fourier expand the fluctuations in the ``time direction'' according to
\be
\ph(\t,x) = \sum_{n=-\infty}^\infty \ph_n(x) e^{2 \pi i n \t /T} \qquad \Rightarrow \qquad \ph_n(x) = \frac{1}{T}\int_0^T d\t \ph(\t,x)e^{-2 \pi i n \t /T}
\ee
with the complex coefficients $\ph_n(x)$ obeying the reality constraint $\ph_n(x) = \ph_{(-n)}(x)^*$. In this case the integration over the momenta in the $\tau$-direction, contained in Tr, is converted into a discrete sum over Matsubara modes. The remainder is given by ${\rm tr}$ which contains the operator trace restricted to the spatial slice together with a trace in field space.\footnote{When performing a ``dimensional reduction'' of a higher-dimensional quantum field theory on a circle, the Matsubara modes are usually called Kaluza-Klein states. A string compactification typically restricts itself to the massless sector. In the language of field theory at finite temperature the latter corresponds to truncating the Matsubara sums considering the zero-mode only.}
\be\label{traceredef}
\begin{split}
{\rm Tr} \rightarrow 
& \,  \sqrt{\epsilon} \, \sum_{n= - \infty}^\infty {\rm tr} \,  .
\end{split}
\ee
The $\t$ derivatives are then converted into Matsubara masses of the fluctuations in time direction
\be
\p_\t^2 \rightarrow - \left( \frac{2 \pi n}{T} \right)^2 \, .
\ee

The expansion of tr in terms of the intrinsic curvature can then be carried out by applying the standard heat-kernel expansion to the Laplace operator on the spatial slices. Applying the trace technology summarized in Appendix \ref{App:B.1} it is then straightforward to find the two lowest terms in the spatial-curvature expansion. The result is most conveniently expressed in terms of the dimensionless analogs of Newton's constant, the cosmological constant and the Matsubara mass $m$
\be\label{dimless}
g_k = G_k k^{d-1} \, , \qquad \lambda_k = \Lambda_k k^{-2} \, , \qquad m = \frac{2 \pi}{Tk} \, .
\ee
The dimensionality of $G_k$ is the canonical one for a $d+1$-dimensional spacetime. In addition, we introduce the anomalous dimension of Newtons constant
\be\label{Zk}
\eta_N \equiv - \p_t \ln (T_k \, Z_{Nk}) \, , \qquad G_k = Z_{Nk}^{-1} \, G_0 \, . 
\ee
Here we anticipated that the size of the time circle may depend on $k$, which would reflect itself through an anomalous running of the Matsubara mass $m \not \sim k^{-1}$.

The contribution of the transverse-traceless modes \eqref{tensortraces} then becomes
\be \label{eq:TgravTT1}
\begin{split}
\cT_\mathrm{TT} = & \, \frac{d_{\rm 2T}   \,  \sqrt{\epsilon} }{(4 \pi)^{d/2}}\, \sum_n \int d^dx \sqrt{\bar{\sigma}} k^d \bigg\{
  q^{1,0}_{d/2}(w_{\rm 2T})  
+  \Rb k^{-2} \left( \tfrac{1}{6} \, q^{1,0}_{d/2-1}(w_{\rm 2T}) - C_{\rm 2T} \, q^{2,0}_{d/2}(w_{\rm 2T})  \right)
\bigg\}\, .
\end{split}
\ee
Here $d_{\rm 2T} \equiv \half (d+1)(d-2)$ arises from the trace over vector indices,
\be\label{def:w2T}
w_{\rm 2T} \equiv \epsilon^{-1} m^2 n^2 - 2 \lambda_k \, , 
\ee
and the $q$-functions are defined in \eqref{qfcts}. The scalar trace \eqref{scalartrace2} can be evaluated along the same lines
\be\label{Sgrav}
\begin{split}
\cT_{{\rm s}} = & \, \frac{\sqrt{\epsilon} }{(4 \pi)^{d/2}}\,  \sum_n \int d^dx \sqrt{\bar{\sigma}} k^d \bigg\{
 q^{1,0}_{d/2}(w_0) + \tfrac{1}{6} \tfrac{\bar{R}}{k^{2}} q^{1,0}_{d/2-1}(w_0) \\
& \, + \tfrac{d-2}{2d\lambda_k} \tfrac{\bar{R}}{k^{2}} \Big[
q^{2,-1}_{d/2}(w_0) - q^{1,0}_{d/2}(w_0) + \left( \tfrac{3}{\epsilon} m^2 n^2 - \tfrac{4(d-3)}{d-2} \lambda_k  \right) q^{2,0}_{d/2}(w_0) \Big]
\bigg\}
\end{split}
\ee
with
\be\label{def:w0}
w_0 \equiv - \tfrac{1}{\lambda_k} \tfrac{d-1}{d-2} \Big( \tfrac{1}{\epsilon} m^2 n^2 - 2 \lambda_k \Big) 
\Big(  \tfrac{1}{\epsilon} m^2 n^2 - \tfrac{d-2}{d-1} \lambda_k \Big) \, . 
\ee

After working out the operator traces on the spatial slices the next step constitutes in performing the Matsubara sums for the ``time-like'' fluctuations. In order to be able to carry out the sums analytically, we specialize to the optimized cutoff \eqref{Ropt}. In this case the threshold functions \eqref{treshold} become particularly simple. Using the results derived in Appendix \ref{App:resum} the transverse-traceless part of the trace can be written as
\be \label{eq:TgravTT2}
\begin{split}
\cT_\mathrm{TT} = & \, \frac{d_{\rm 2T}   \,  \sqrt{\epsilon} }{(4 \pi)^{d/2}}   \int d^dx \sqrt{\bar{\sigma}} k^d \bigg\{
  T^{1,0}_{d/2} 
+  \tfrac{\bar{R}}{k^{2}} \left( \tfrac{1}{6} \, T^{1,0}_{d/2-1} - C_{\rm 2T} \, T^{2,0}_{d/2}  \right)
\bigg\}
\end{split}
\ee
while the scalars contribute
\be\label{Sgrav2}
\begin{split}
\cT_{{\rm s}} = & \, \frac{\sqrt{\epsilon}}{(4 \pi)^{d/2}}  \int d^dx \sqrt{\bar{\sigma}} k^d \bigg\{
 S^{1,0}_{d/2}  + \tfrac{1}{6} \tfrac{\bar{R}}{k^{2}} S^{1,0}_{d/2-1} 
+ \tfrac{d-2}{2d\lambda_k} \tfrac{\bar{R}}{k^{2}} \Big[
\left( 1 - \tfrac{4(d-3)}{d-2} \lambda_k  \right) \, S^{2,0}_{d/2} - S^{1,0}_{d/2} +  3 S^{2,1}_{d/2} \Big]
\bigg\} \, . 
\end{split}
\ee
The functions $T^{p,q}_{l}$ and $S^{p,q}_{l}$ are defined in \eqref{sumfcts} and their explicit forms are obtained in Appendix \ref{App:resum}.

The non-perturbative beta functions for the Newton constant and the cosmological constant, arising from the truncation \eqref{eq:decomposedEHaction}, are encoded in the coefficients of the zeroth and first order terms of the intrinsic curvature. Taking the $t$ derivative and setting $\chi = \bar{\chi}$ afterward, the l.h.s.\ of the flow equation becomes
\be
\p_t \Gamma_k \big|_{\chi = \bar{\chi}} = \frac{\sqrt{\epsilon}}{16 \pi} \int d^dx \sqrt{\bs} \left\{ - \Rb \, \p_t \left( \tfrac{T_k}{G_k}\right) + 2 \, \p_t \left( \tfrac{\Lambda_k}{G_k} T_k \right) \right\} \,  .
\ee
Equating these curvature terms with the corresponding terms in \eqref{eq:TgravTT2} and \eqref{Sgrav2} then gives rise to the beta functions for the dimensionfull Newton constant and cosmological constant. In terms of the dimensionless quantities \eqref{dimless} these read
\be \label{eq:betaFct}
\p_t g_k = \beta_g(g, \lambda; m) \, , \qquad \p_t \lambda_k = \beta_\lambda(g, \lambda; m)
\ee
where
\be\label{beta1}
\begin{split}
\beta_g = & \, \left( d-1 + \eta_N \right) \, g \, , \\
\beta_\lambda = & \, (\eta_N - 2 ) \lambda + \frac{4 \, m \, g}{(4 \pi)^{d/2}}  \left( d_{\rm 2T} \, T^{1,0}_{d/2} + S^{1,0}_{d/2} \right) \, . 
\end{split}
\ee
The anomalous dimension of Newton's constant is given by
\be
\eta_N = \frac{m \, g \, B_1(\lambda) }{1 + m \, g \, B_2(\lambda)}
\ee
with
\be \label{eq:B1B2}
\begin{split}
B_1 = & \, \frac{8}{(4 \pi)^{d/2}} \bigg\{
d_{\rm 2T} \left( \tfrac{1}{6} \Upsilon^{1,0}_{d/2-1} - C_{\rm 2T} \Upsilon^{2,0}_{d/2} \right) + \tfrac{1}{6} \Psi^{1,0}_{d/2-1} \\
& \qquad \qquad + \tfrac{d-2}{2d\lambda} \left( \left( 1 - \tfrac{4(d-3)}{d-2} \lambda \right) \Psi^{2,0}_{d/2} - \Psi^{1,0}_{d/2} + 3  \Psi^{2,1}_{d/2} \right)
\bigg\} \, , \\
B_2 = & \, \frac{4}{(4 \pi)^{d/2}}  \bigg\{
d_{\rm 2T} \left( \tfrac{1}{6} \tilde{\Upsilon}^{1,0}_{d/2-1} - C_{\rm 2T} \tilde{\Upsilon}^{2,0}_{d/2} \right) + \tfrac{1}{6} \tilde \Psi^{1,0}_{d/2-1} \\
& \qquad \qquad +   \tfrac{d-2}{2d\lambda} \left( \left( 1 - \tfrac{4(d-3)}{d-2} \lambda \right) \tilde \Psi^{2,0}_{d/2} - \tilde \Psi^{1,0}_{d/2} + 3  \tilde \Psi^{2,1}_{d/2} \right)
\bigg\} \,  . \\
\end{split}
\ee
The beta functions \eqref{eq:betaFct} constitute the final result of this section. Besides the dimensionless Newton constant and cosmological constant they parametrically depend on the Matsubara mass $m$, which encodes the size of the foliation in units of the IR-scale $k$. 

\section{RG flow scenarios}
\label{sect.5}
In this section we proceed by analyzing the RG flow emanating from the beta functions \eqref{eq:betaFct}. Our prime interest thereby lies on the influence of signature effects on the existence of non-trivial fixed points of the RG flow which could feature in a gravitational Asymptotic Safety scenario. Concretely, the general properties of the beta functions will be discussed in Sect.\ \ref{sect:5.1}. Their fixed-point structure in the limits $m \rightarrow \infty$ and $m \rightarrow 0$ are detailed in Sects.\ \ref{sect:5.2} and \ref{sect:5.3}, respectively, while the floating fixed point scenario with finite $m$ is elucidated in Sect.\ \ref{sect:5.4}.

\subsection{Analytic structure of the beta functions}
\label{sect:5.1}
The key feature for understanding the RG flow implied by \eqref{eq:betaFct} is the analytic structure of the beta functions \eqref{beta1}. Eq.\ \eqref{eq:mastersum} illustrates that summing up the tower of Matsubara states either leads to hyperbolic functions which are well-defined on the entire real axis or trigonometric functions which give rise to poles for certain values $x$. Which situation is actually realized is determined by the relative sign between the kinetic ($\p_\tau^2$)-terms and the potential terms including the spatial momenta and cosmological constant.

In an Euclidean theory with conventional (healthy) kinetic terms the $n^2$ and $x^2$ terms come with a relative plus sign, reflecting that the ``time''-like and spatial momenta stem from a positive semi-definite Laplace operator. In this case the Matsubara sums solely give rise to hyperbolic terms. Performing the Wick rotation to Lorentzian signature changes the relative sign and thus one encounters trigonometric terms. Essentially, this reflects the fact that in the two settings the symbol of the differential operator appearing in the definition of the kinetic term is positive definite or of mixed signature.

For the truncation \eqref{eq:decomposedEHaction} this simple picture does not hold. Inspecting the threshold functions entering into \eqref{beta1}, a careful analysis of the signs of the roots \eqref{wpm} indicates that the analytic structure of the beta functions depends on the value of the dimensionless cosmological constant $\lambda$. As indicated in Table \ref{tab:analyticStructure} there are three distinguished regions in the $g$-$\lambda$ plane.   
\begin{table}[t!]
\begin{center}
\begin{tabular}{|c||c|c|c|}
 \hline
 $\epsilon$ & $\lambda < - 4 C_0 $ & $-4 C_0 < \lambda < 1/2$ & $1/2 < \lambda $ \\
 \hline
 $+1$ & hyperbolic & mixture & trigonometric \\
 \hline
 $-1$ & trigonometric & mixture & hyperbolic \\
 \hline 
\end{tabular}
\caption{The analytic structure of the beta functions \eqref{eq:betaFct} depends on the signature of spacetime $\epsilon$, the value of the cosmological constant, and the dimension of the spatial slice via the combination $C_0$ \eqref{eq:abbreviations}. We distinguish the cases where the threshold functions entering the beta functions are build from hyperbolic or trigonometric functions only or contain both types of terms.} \label{tab:analyticStructure}
\end{center}
\end{table}
For $\lambda < - 4 C_0$ the beta functions for Euclidean and Lorentzian signature are build only from hyperbolic and trigonometric functions, respectively. For $\lambda > 1/2$ the analytic properties are interchanged, i.e., the Euclidean signature leads to trigonometric terms while the beta functions originating from Lorentzian signature are purely hyperbolic. In the intermediate region one always encounters a mix of hyperbolic and trigonometric terms. 

The latter property can be traced back to the structure of the functional traces \eqref{fl2} contributing to the flow. For Euclidean signature, the transverse-traceless fluctuations give rise to hyperbolic terms while the conformal mode contributes the trigonometric pieces. Ultimately, the occurrence of the trigonometric terms is a consequence of the well-known conformal factor problem of gravity:\footnote{The conformal factor problem is also present in the metric formulation of the FRGE. In this framework the wrong-sign kinetic term of the $h$ field is compensated by a suitable choice of regulator function $\cR_k$. The structure of the flow equation \eqref{FRGE} then ensures that the conformal mode contributes to the flow exactly like a scalar field with positive-definite kinetic term.} Performing the second variation of the Ricci scalar leads to ``wrong-sign'' spatial momenta for the conformal scalar $h$. Combined with the standard ``time'' derivatives of the ansatz \eqref{eq:decomposedEHaction} this results in a ``wrong'' relative sign between kinetic and potential contributions. In other words the {\it Euclidean conformal sector} essentially contributes as a standard scalar field in a {\it Lorentzian signature spacetime}. As a consequence, the beta functions develop poles at values $x$, where a given Matsubara mode leads to a vanishing denominator in \eqref{eq:mastersum}. 

\subsection{The compactification limit}
\label{sect:5.2}
We now study the flow of the beta functions \eqref{eq:betaFct} for $k$-independent $T$ in the limit of a collapsing time circle, $T \rightarrow 0$. In this limit, the Matsubara modes become infinitely heavy and decouple, so that the flow becomes essentially $d$-dimensional. In order to account for this effect we trade the $D$-dimensional Newton constant $G_N^{(D)}$ appearing in the Einstein-Hilbert ansatz \eqref{eq:decomposedEHaction} for its $d$-dimensional analog
\be\label{dimreduc}
G_k^{(d)} \equiv G_k^{(D)} \, T^{-1} \, . 
\ee
At the level of the dimensionless couplings \eqref{dimless}, this relation translates into
\be\label{gddef}
m \, g^{(D)}_k = 2 \pi \, g^{(d)}_k \, . 
\ee

The beta functions governing the scale dependence of  $g^{(d)}_k$ can then be deduced by first rewriting \eqref{eq:betaFct} in terms of the $d$-dimensional Newton constant and subsequently taking the limit $m \rightarrow \infty$. Utilizing the intermediate results of Appendix \ref{App:C.1} we obtain
\be \label{flowred}
\p_t g_k^{(d)} = \beta_g^{(d)}(g^{(d)}, \lambda) \, , \qquad \p_t \lambda_k = \beta_\lambda(g^{(d)}, \lambda)
\ee
where
\be\label{betared}
\begin{split}
\beta_g^{(d)} = & \, \Big( d-2 + \eta_N^{(d)} \Big) \, g^{(d)} \, , \\
\beta_\lambda = & \, \Big( \eta_N^{(d)} - 2 \Big) \lambda + \frac{2\, g^{(d)}}{(4 \pi)^{d/2-1}}  \Big( d_{\rm 2T} +1 \Big) \Big( \Phi_{d/2}^{1,0}(-2\lambda) - \half \eta_N^{(d)} \tilde{\Phi}_{d/2}^{1,0}(-2\lambda) \Big) \, . 
\end{split}
\ee
The anomalous dimension of the $d$-dimensional Newton constant is given by
\be\label{adimd}
\eta_N^{(d)} = \frac{2 \pi g^{(d)} \, B_1(\lambda) }{1 + 2 \pi g^{(d)} \, B_2(\lambda)}
\ee
where the functions $B_1$ and $B_2$ have simplified to
\be
\begin{split}
B_1 =  \tfrac{8}{(4 \pi)^{d/2}}  \bigg( & \, \tfrac{1}{6} \, (d_{\rm 2T} + 1) \, \Phi^{1,0}_{d/2-1} - \tfrac{d-2}{2d\lambda} \Phi^{1,0}_{d/2}
+ \big(\tfrac{d-2}{2d\lambda} - 
d_{\rm 2T} C_{\rm 2T} - \tfrac{2(d-3)}{d} 
\big) \Phi^{2,0}_{d/2}\bigg) \, , \\
B_2 =  \tfrac{4}{(4 \pi)^{d/2}}  \bigg( & \, \tfrac{1}{6} \, (d_{\rm 2T} + 1)  \, \tilde{\Phi}^{1,0}_{d/2-1} - \tfrac{d-2}{2d\lambda} \tilde{\Phi}^{1,0}_{d/2} 
 + \big(\tfrac{d-2}{2d\lambda} - 
d_{\rm 2T} C_{\rm 2T} - \tfrac{2(d-3)}{d} 
\big) \tilde{\Phi}^{2,0}_{d/2} \bigg)
\end{split}
\ee
and all threshold functions are evaluated at $w = -2\lambda_k$.

At this stage the following remarks are in order. Inspecting \eqref{betared} we first observe that the part of the beta functions induced by the mass dimension of $G_k^{(d)}$ matches the one expected for the Newton constant in $d$ dimensions. Moreover, the signature dependence of the beta functions has dropped out in the compactification limit. This is expected, since all information on the signature of spacetime has been encoded in the $T$ circle. Finally we note that the quantum contributions to the running of the cosmological constant and the Newton constant are essentially provided by the transverse-traceless fluctuations on the $d$-dimensional spatial slice and one scalar field. This is precisely the off-shell-field content of $d$-dimensional gravity upon gauge fixing diffeomorphism invariance. In this light, the decoupling of the transverse spatial vector observed in \eqref{cutoffs} seems necessary to reproduce the correct $T \rightarrow 0$ limit of the gravitational flow equations. 
\begin{table}[t!]
\begin{center}
\begin{tabular}{|c||c|c||c|c|}
 \hline
\; $d$ \; & $g^{(d)}_*$ & $\lambda_*$ & $\tau_*$ & $\theta_{1,2}$ \\ \hline
 $3$ & \; $ 0.24 $ \; & \; $ 0.30 $ \; & \; $0.02$ \; & \; $ 0.89 \pm 3.22 i $ \; \\
 $4$ &  $0.77 $ & $0.28 $ & $0.22 $ & $2.69 \pm  4.63 i$ \\ 
 $5$ & $3.17 $ & $0.29 $ & $0.62 $ & $4.55 \pm  6.26 i $ \\
 $6$ & $15.3 $ & $0.29 $ & $1.14 $ & $6.64 \pm  7.86 i$ \\
 $7$ & $84.0 $ & $0.30 $ & $1.75 $ & $8.97 \pm 9.41 i$ \\ \hline
\end{tabular}
\caption{Position and critical exponents of the non-Gaussian fixed point of the beta functions \eqref{betared} for selected spatial dimensions $3 \le d \le 7$. The universal product $\tau_*$, eq.\ \eqref{univprod}, and the critical exponents of the NGFP are very similar to the ones found in the metric Einstein-Hilbert truncation \cite{Fischer:2006fz,Fischer:2006at}.} \label{redFPs}
\end{center}
\end{table}

A crucial ingredient in understanding the RG flow implied by the system \eqref{flowred} are the fixed points of the RG flow which by definition satisfy $\beta_g(g^{(d)}_*, \lambda_*) = 0$, $\beta_\lambda(g^{(d)}_*, \lambda_*) = 0$. The beta functions posses a Gaussian fixed point (GFP) $g^{(d)}_* = 0, \lambda_* = 0$. This fixed point corresponds to the free or Gaussian theory. Moreover, the beta functions also give rise to a unique non-Gaussian fixed point (NGFP) with $g^{(d)}_* > 0, \lambda_* \not = 0$ for any dimension $d > 3$ whose position can easily be found numerically. For $3 \le d \le 7$ their values together with the universal scaling variable
\be\label{univprod}
 \tau_* \equiv  \lambda_* \, \left( g_*^{(d)} \right)^{2/(d-2)}  
\ee
are summarized in Table \ref{redFPs}. Given a fixed point of the flow equation, the linearized flow near the fixed point is governed by the Jacobi matrix ${\bf B} = (B_{ij}) \equiv \p_j \beta_i|_{g^{(d)} = g^{(d)}_*, \lambda = \lambda_* }$. The stability properties of the fixed point are then encoded in the stability parameters $\theta$ given by minus the eigenvalues of  $\bf B$. For the NGFP of our system these are given in the last column of Table \ref{redFPs}. The positive real part of $\theta$ thereby indicates that the NGFP is UV-attractive in both $g_*^{(d)}$ and $\lambda_*$ while the non-zero imaginary part implies that RG trajectories coming close to the fixed point in the UV will spiral into the NGFP as $k \rightarrow \infty$.

Remarkably, the properties of the NGFP reported in Table \ref{redFPs} are very similar to the NGFP featuring in the Asymptotic Safety program for metric gravity. In particular the universal scaling variable $\tau_*$ and the critical exponents turn out to be similar to the typical values obtained in the analysis of the beta functions of the Einstein-Hilbert action in the metric formalism \cite{Fischer:2006fz,Fischer:2006at}. Thus it is tempting to speculate that the UV behavior of quantum gravity in the metric formulation and the ADM-decomposed formulation at $T = 0$ are actually governed by the same universality class. 
\begin{figure}[t!]
\begin{center}
\includegraphics[width=0.47\textwidth]{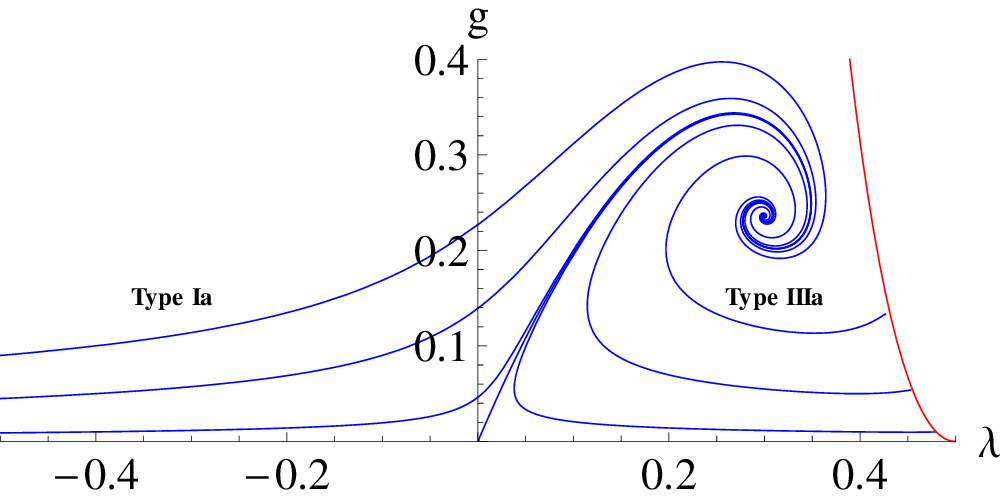}
\includegraphics[width=0.47\textwidth]{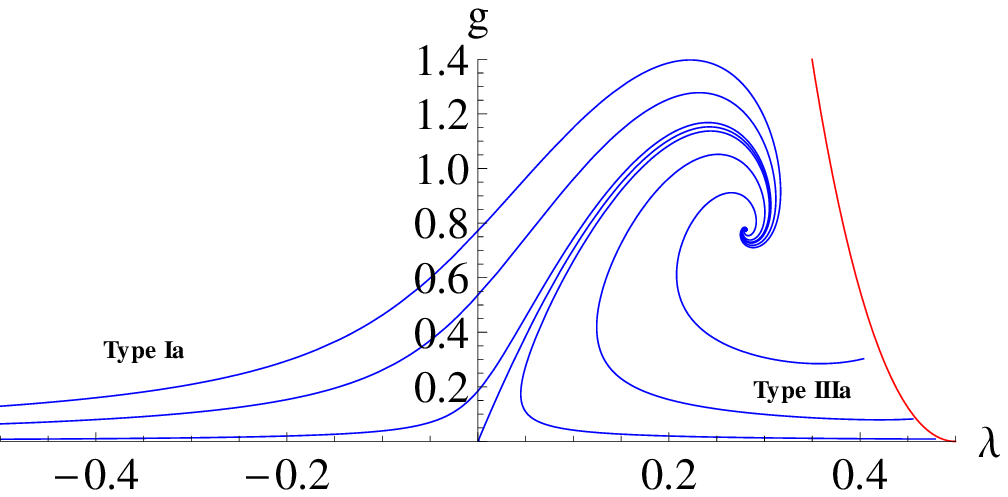}
\caption{Phase diagram originating from the foliated Einstein-Hilbert truncation in the limit $T \rightarrow 0$ in $d=3$ (left) and $d=4$ (right).  Along the red line the anomalous dimension \eqref{adimd} diverges. 
The flow is strikingly similar to the one obtained within the metric Einstein-Hilbert truncation \cite{Reuter:2001ag}.}
\label{fig:compactificationLimitFlow}
\end{center}
\end{figure}

The phase diagram obtained by integrating the beta functions \eqref{betared} numerically is shown in Fig. \ref{fig:compactificationLimitFlow}. The UV-behavior of all trajectories with positive Newton constant is governed by the NGFP. Tracing the RG flow towards the IR, there are three classes of trajectories: the separatrix (Type IIa) connects the NGFP with the GFP. Trajectories which pass to its left (right) are classified as Type Ia (Type IIIa), respectively, and lead to a negative (positive) IR-value of the cosmological constant. The flow pattern is virtually identical to the one obtained in the metric Einstein-Hilbert truncation \cite{Reuter:2001ag}.

\subsection{The decompactification limit}
\label{sect:5.3}
In the last subsection, we found that in the limit $T \rightarrow 0$ all signature effects drop out of the beta functions \eqref{beta1}. In this subsection we analyze the opposite limit where the foliation becomes infinitely extended $T \rightarrow \infty$. In this limit, the discrete Matsubara sums in \eqref{eq:mastersum} become continuous integrals. As a consequence, the trigonometric terms in the beta functions develop a branch-cut singularity leading to diverging expressions. Thus the beta functions make sense in the case where they consist of hyperbolic terms only. Inspecting Table \ref{tab:analyticStructure} shows that the beta functions are well-defined in the regions $\lambda < -4C_0$ in Euclidean and $\lambda > 1/2$ for Lorentzian signature only. In particular they become ill-defined in the central region $- 4 C_0 < \lambda < 1/2$ for both signatures, since there they always include both hyperbolic and trigonometric terms.

From Table \ref{redFPs} we note that the NGFP in the compactification limit is situated in the central region. As a consequence we do not expect to find a NGFP for the $T \rightarrow \infty$ limit. Indeed a numerical analysis confirms that there is no NGFP for positive Newton constant. This should, however, not be interpreted as a failure of Asymptotic Safety. As discussed in Subsection \ref{sect:5.1} the appearance of the trigonometric terms and thus the break down of the beta functions in the central region originates from the conformal-factor problem intrinsic to the ansatz \eqref{eq:decomposedEHaction}. Clearly, our  truncation is too simple to allow for a dynamical resolution of this problem. Addressing this problem systematically presumably requires the inclusion of higher derivative operators like the square of the intrinsic Ricci scalar $({}^{(d)}R)^2$ in the truncation ansatz. Since the inclusion of such higher-derivative terms in the truncation ansatz is known to be very laborious \cite{oliver3,Lauscher:2002sq,Rechenberger:2012pm} and beyond a first investigation of foliated RG flows, we do not pursue this direction further and leave it to future research.

\subsection{RG flows at finite $m$}
\label{sect:5.4}
In our final analysis we complement the system \eqref{eq:betaFct} by a beta function for the Matsubara mass $m_k$. Our basic assumption is that the dimensionless $m_k$ also approaches a finite value $m_*$ at the NGFP. The supplementary beta function, capturing this behavior is
\be\label{betam}
\p_t m_k = \beta_m(g, \lambda, m)
\ee
where the NGFP is a common zero of all three beta functions
\be
\beta_g(g_*, \lambda_*, m_*) = 0 \, , \quad \beta_\lambda(g_*, \lambda_*, m_*) = 0 \, , \quad \beta_m(g_*, \lambda_*, m_*) = 0
\ee
with $m_* \not = 0$. Since the computation of $\beta_m$ is beyond the scope of the present work, we will approximate $m_k = m_* \not = 0$ as a free parameter and investigate the parametric dependence of \eqref{eq:betaFct} on $m = m_*$.

\begin{figure}[t!]
\begin{center}
\includegraphics[width=0.45\textwidth]{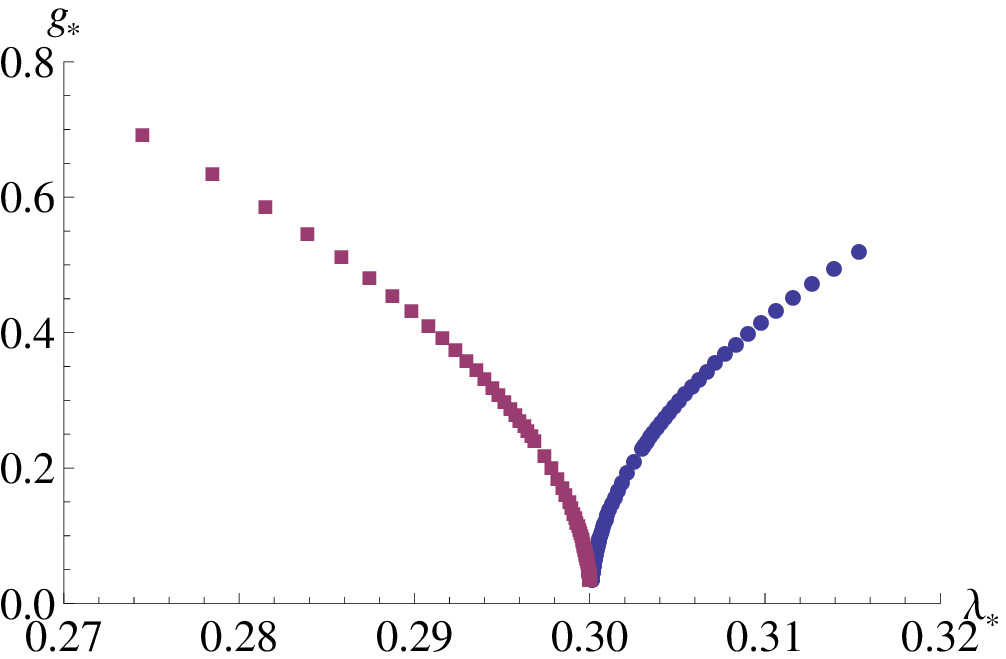} \, 
\includegraphics[width=0.45\textwidth]{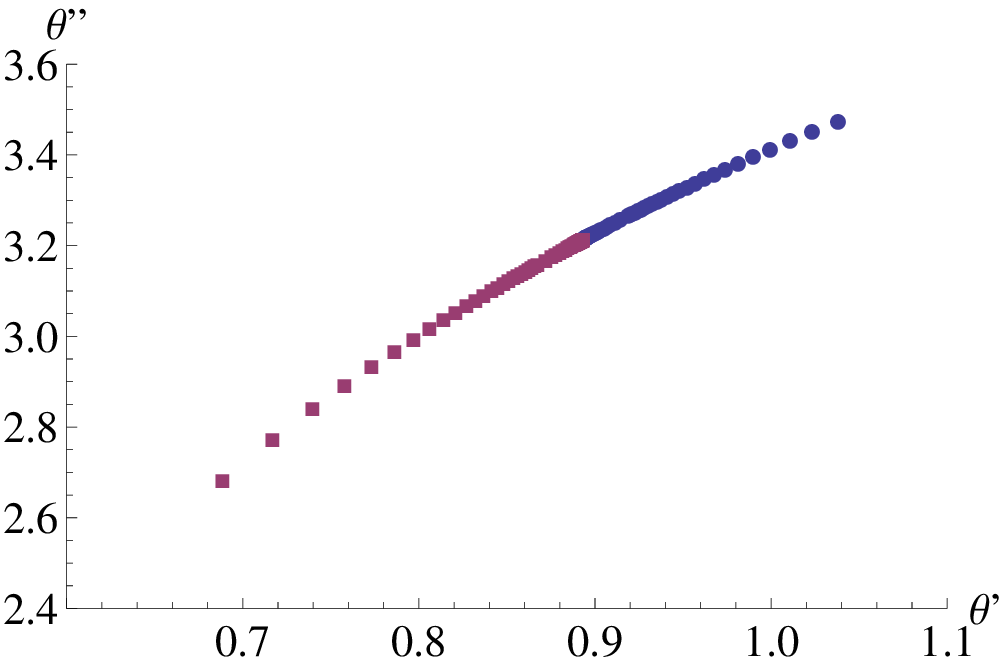} \\[2.5ex]
\includegraphics[width=0.45\textwidth]{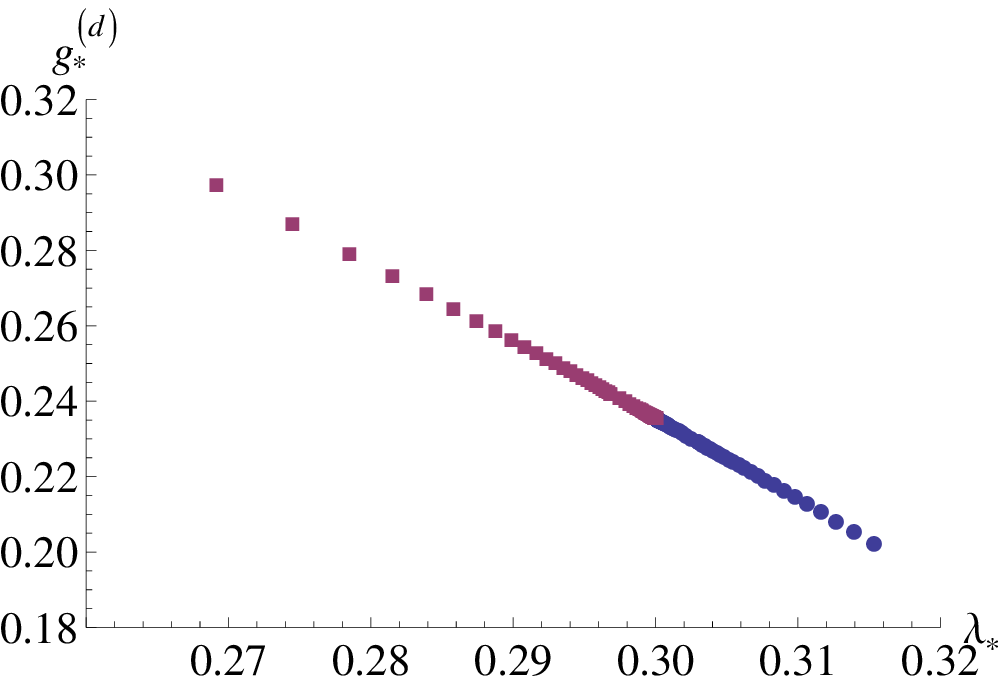}
\caption{Parametric $m$ dependence of the position $\{g_*, \lambda_*\}$ of the NGFP (upper left panel) and its complex stability coefficients $\theta_{1,2} = \theta' \pm i \theta''$ (upper right panel). The corresponding location in terms of the $d$-dimensional Newton constant \eqref{gddef} is shown in the lower panel. The blue circles and magenta squares correspond to Euclidean and Lorentzian signature respectively. In the compactification limit $m\rightarrow \infty$ the Lorentzian and Euclidean NGFP coincide.}
\label{NGFPmdep}
\end{center}
\end{figure}
An immediate consequence of approximating $m = const, \p_t m_k = 0$ is that the dimensionfull length of the foliation direction scales with $k$ according to its canonical dimension $T \propto k^{-1}$. In \cite{Manrique:2011jc} this was dubbed the ``floating fixed point scenario''. In this scenario the time-interval length is inverse proportional to the RG scale $k$. This means that the time circle decompactifies in the IR as one would expect from a physically viable theory. On the other hand in the deep UV the circle collapses in a controlled way, so that the presence of the ``extra dimension'' and signature still enters into the beta function through the dimensionless parameter $m$.

The possibility of this additional $k$ dependence has already been anticipated in the definition of the anomalous dimension of the Newton constant, \eqref{Zk}, so that the beta functions \eqref{beta1} remain valid in this case as well. For $m > \sqrt{\frac{2(4d-3)}{d-1}}$ the beta function for both signatures remain regular throughout the entire central region. When giving numerical results we will choose $m = 2\pi$ for illustration purposes.

We analyze the $m$ dependence of the NGFP arising from the system \eqref{beta1} in $d=3$, corresponding to a $D=3+1$-dimensional spacetime. In Subsection \ref{sect:5.2} we established the existence of the NGFP in the limit $m \rightarrow \infty$. In this case the Euclidean and Lorentzian NGFP fall on top of each other. In terms of the dimensionless Newton constant in $D$ dimensions $g_*^{(D)} \equiv g_*$ it is situated at $\{g_*^{(D)}, \lambda_*\} = \{ 0, 0.3 \}$. The parametric $m$ dependence of the NGFP is shown in Fig.\ \ref{NGFPmdep}.
Lowering $m$, the signature effects drive the Euclidean and Lorentzian NGFP apart. However, the signature dependence is rather mild, so that its location and critical exponents are rather robust with respect to a change in $m$. In particular we observe that the real part of the critical exponents is always positive so that the NGFP is UV attractive in both $g_*$ and $\lambda_*$ for all values $m$. Thus the NGFP constitutes a viable candidate fixed point for the gravitational Asymptotic Safety program in both Euclidean and Lorentzian signature spacetimes. 

We illustrate these findings by giving the numerical results for the position and critical exponents of the NGFP at $m = 2 \pi$ in Table \ref{tab:fixedPoints}.
\begin{table}[t!]
\begin{center}
\begin{tabular}{|c||c|c||c|c|}
 \hline
 $\epsilon$ & $g_*$ & $\lambda_*$ & $g_* \lambda_*$ & $\theta_{1,2}$ \\ \hline
 $+1$ &  $0.23$ &  $0.30$ &  $0.070$  &  $0.92 \pm 3.27 i$ \\
 $-1$ &  $0.24$ &  $0.30$ &  $0.072$  &  $0.87 \pm 3.16 i$ \\ \hline 
\end{tabular}
\caption{Fixed point values and critical exponents in the floating fixed point scenario for $m = 2\pi$ and $D=3+1$.} \label{tab:fixedPoints}
\end{center}
\end{table}
In this case the signature affects the value of the universal product $g_* \lambda_*$ and the real and imaginary part of the critical exponents at the $5\%$ level. This highlights the remarkable robustness of the flow implied by \eqref{eq:betaFct} with respect to changing the spacetime signature.

\begin{figure}[t!]
\begin{center}
\includegraphics[width=0.45\textwidth]{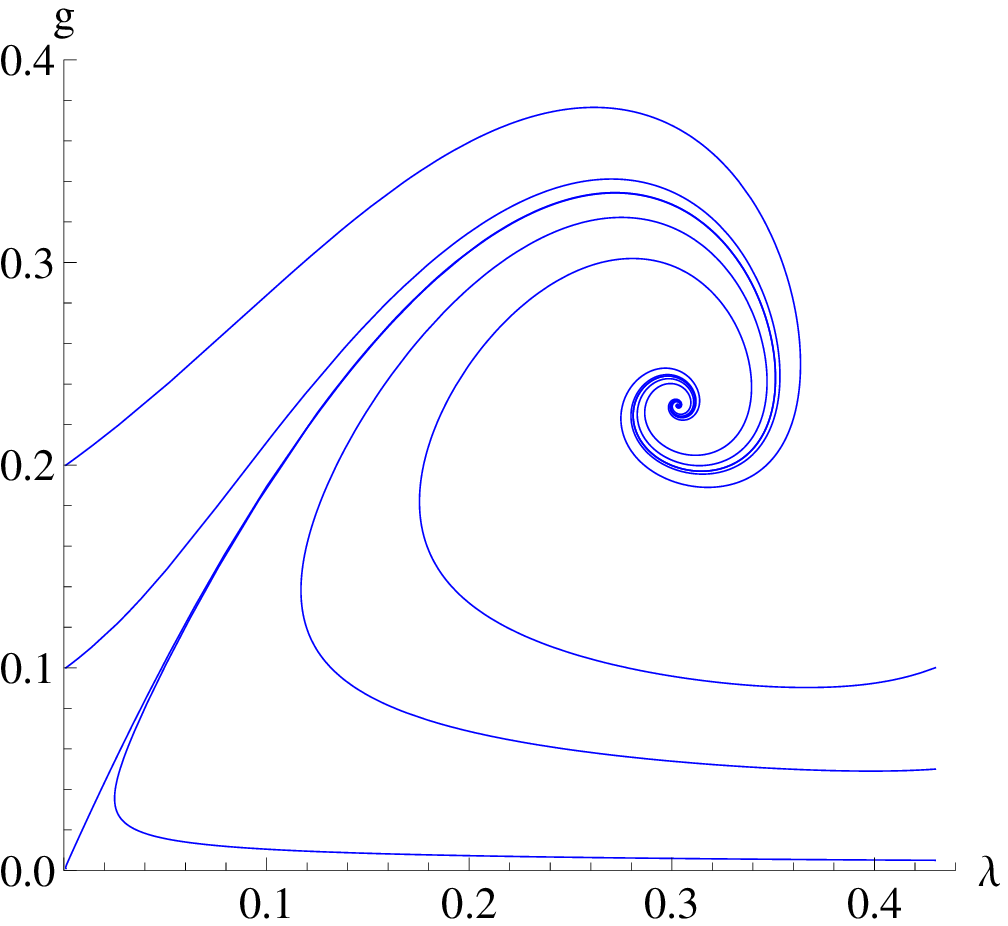} \;
\includegraphics[width=0.45\textwidth]{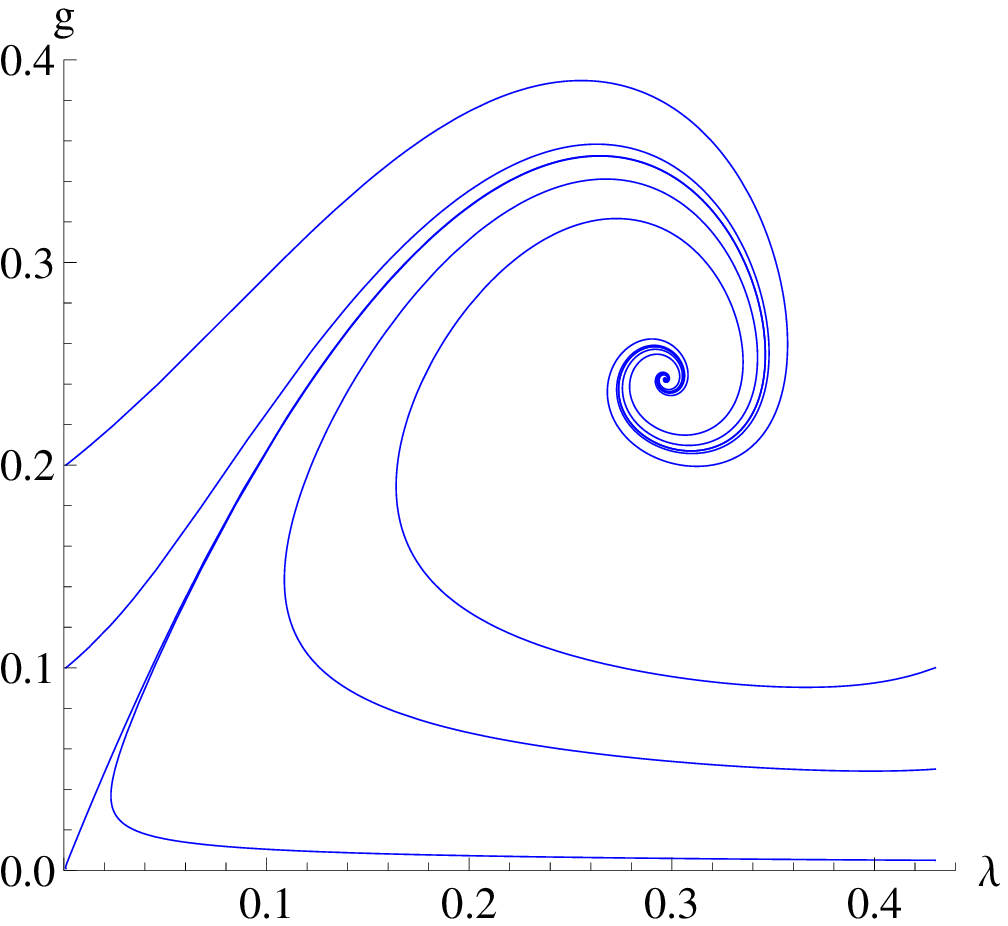}
\caption{Flow diagram for Euclidean (left panel) and Lorentzian (right panel) signature for $m = 2\pi$ in $D=3+1$ spacetime dimensions.}
\label{fig:m2piFlow}
\end{center}
\end{figure}

This rather remarkable stability also extends away from the NGFP. This feature is illustrated in Fig.\ \ref{fig:m2piFlow} where we constructed the phase diagrams for Euclidean and Lorentzian signature flows by numerically integrating the flow equations for the $D$-dimensional Newton constant $g_k$ and the cosmological constant $\lambda_k$ for $m = 2\pi$. In both cases the flow is dominated by the interplay between the NGFP in the UV and the Gaussian fixed point located at the origin $\{g_*, \lambda_*\} = \{ 0,0\}$ in the infrared. The flow diagrams turn out virtually identical, consolidating the assessment that the signature of spacetime plays a subleading role when determining the gravitational RG flow. Moreover, they coincide with their counterparts obtained in the compactification limit displayed in Fig.\ \ref{fig:compactificationLimitFlow}. Furthermore the corresponding classification of RG trajectories is the same as the one obtained from the flow equations in the metric formulation \cite{Reuter:2001ag}.
%

\section{Conclusions and outlook}
\label{sect.6}
In this work we constructed the Wetterich equation \cite{Wetterich:1992yh} which captures the non-perturbative RG flow of the effective average action $\Gamma_k$ on the theory space of projective Ho\v{r}ava-Lifshitz gravity \cite{Horava:2008ih}. Its solutions interpolate continuously between the classical action $S=\G_{k\rightarrow\infty}$ and the full quantum effective action $\G=\G_{k\rightarrow 0}$. In contrast to the metric construction \cite{Reuter:1996cp}, this novel functional renormalization group equation (FRGE) encodes the gravitational degrees of freedom in the lapse function, the shift vector and the spatial metric originating from the ADM decomposition \cite{Arnowitt:1959ah}. The construction imprints a foliation structure on the quantum spacetime, giving rise to a Euclidean ``time'' direction. Owed to the foliation, the flow equation closely resembles the ones encoding the RG flows of quantum field theory at finite temperature \cite{Litim:2006ag}.

Analogously to the metric construction \cite{Reuter:1996cp}, our FRGE for the ADM fields employs the background field method with a linear background split and constructs the IR regulator from the background quantities. The resulting RG flows explicitly preserve background-foliation-preserving symmetries Diff($\cM, \Sigma$) $\subset $ Diff($\cM$). This invariance ensures that the RG flow does not generate interactions that violate this symmetry dynamically. Thus our equation constitutes a natural starting point for studying the RG flow of anisotropic gravity theories like Ho\v{r}ava-Lifshitz gravity, possibly shedding light on hitherto open questions concerning the perturbative renormalizability and a possible dynamical restoration of Lorentz symmetry at low energies of the model.

As explained in Sect.\ \ref{sect.3} there is a clear relation between the RG flows captured by the metric formulation of gravity and projective Ho\v{r}ava-Lifshitz gravity. While, upon gauge-fixing, the off-shell field content of the two theories looks identical, the weaker symmetry requirements of the anisotropic case entail that the theory space of Ho\v{r}ava-Lifshitz gravity contains the one of metric gravity as a subspace. The result obtained in \cite{Reuter:1996cp}, showing that metric formulation of the FRGE does not generate non-diffeomorphism invariant interactions dynamically, indicates that the latter is closed under RG transformations.

As a first application of our novel FRGE, we studied the RG flow captured by the ADM-decomposed Einstein-Hilbert action \eqref{eq:decomposedEHaction}. The beta functions encode the scale dependence of the Newton constant and the cosmological constant and parametrically depend on the dimensionless Matsubara mass $m$, associated with the ``size'' of the ``time direction''. As a key result, already anticipated in \cite{Manrique:2011jc}, we establish that the RG flow resulting from the foliated flow equation gives rise to a non-Gaussian fixed point intrinsic to the gravitational Asymptotic Safety scenario. This result is independent of the signature of spacetime, as long as $m$ is sufficiently large. Moreover, the phase portraits shown in Figs.\ \ref{fig:compactificationLimitFlow} and \ref{fig:m2piFlow} obtained from the foliated flow equation are strikingly similar to the ones found within the metric Einstein-Hilbert truncation \cite{Reuter:2001ag}. Based on this result, we expect that the foliation structure of spacetime should not play a crucial role when determining the short distance behavior of the theory. 

Naturally, there are a whole range of applications that can be explored based on the FRGE \eqref{eq:wetterichEquation}. Obviously, expressing the gravitational degrees of freedom in terms of the ADM fields together with the invariance under background-foliation-preserving diffeomorphisms prepares the stage for studying the RG flow of Ho\v{r}ava-Lifshitz gravity. The truncation ansatz \eqref{eq:decomposedEHaction} can straightforwardly by supplemented by interaction monomials which break Diff($\cM$) but are invariant under Diff($\cM, \Sigma$). Interesting examples could include a higher-derivative term in the intrinsic curvature $\int d^2x dt \, \sqrt{\sigma} N  \left[ {}^{(d)}R \right]^2$ and the square of the Cotton tensor $\int d^3x dt \, \sqrt{\sigma} N C_{ij} C^{ij}$ which are at the heart of the perturbative renormalizability of the theory. Moreover, the inclusion of matter fields with anisotropic couplings is straightforward. The simplest extension which includes a scale-dependent coupling in the extrinsic curvature terms is currently under investigation and the results will be reported elsewhere \cite{inprep}.

The presence of the foliation structure also constitutes an important step toward connecting the continuum FRGE to the Monte-Carlo simulations of the gravitational partition function within the Causal Dynamical Triangulations (CDT) program \cite{Ambjorn:2012jv}. In a sense our novel flow equation contains the continuum analogue of the foliation structure that is supposed to be responsible for the well-defined classical limit in CDT. Following the ideas \cite{oliverfrac,Horavafit,Reuter:2011ah,Rechenberger:2012pm} one could, e.g., use the foliated flow equation to study diffusion processes on spatial slices and examine if these can be matched to the CDT data recently reported in \cite{Ambjorn:2010fv}. This may shed new light on the question whether the continuum limit of CDT corresponds to an isotropic or anisotropic gravity theory as suggested in \cite{Ambjorn:2010hu,Anderson:2011bj}.

\section*{Acknowledgments}
We thank R.\ and N.\ Alkofer, A.\ Contillo, J.\ Laiho, M.\ Reuter, and O.\ Zanusso for illuminating discussions and E.\ Manrique for participation in the early stages of the project.  The research of S.\ R.\ and F.\ S.\ is supported by by the Deutsche Forschungsgemeinschaft (DFG) within the Emmy-Noether program (Grant SA/1975 1-1).

\begin{appendix}

\section{Trace technology}
\label{App:B}
In this appendix, we collect the technology and intermediate results used for evaluating the functional traces \eqref{tensortraces} and \eqref{scalartraces}. We start with the collection of some standard heat-kernel techniques for evaluating the operator traces on the spatial slices in subsection \ref{App:B.1} while the Matsubara sums are discussed in subsection \ref{App:resum}.

\subsection{Traces on the spatial slice}
\label{App:B.1}
The right hand side of the flow equation \eqref{fl2} naturally gives rise to operator traces ${\rm tr} [ W(\Delta)]$, restricted to the spatial slices and containing functions of the spatial Laplacian $\Delta = - \bs^{ij} \bar{D}_i \bar{D}_j$. Introducing the Laplace anti-transform $\widetilde{W}(s)$
\be
W(z) = \int_0^\infty ds \, \widetilde{W}(s) \, e^{-z s}
\ee
these operator traces can be written as
\be\label{trafo1}
{\rm tr}\left[ W(\Delta) \right] = \int_0^\infty ds \, \widetilde{W}(s) \, {\rm tr} \left[ e^{-s \Delta} \right] \, . 
\ee
The trace of the heat kernel has the well-known asymptotic expansion for $s \rightarrow 0$
\be\label{heatkernel}
{\rm tr} \left[ e^{-s \Delta} \right] = \frac{1}{(4 \pi s)^{d/2}} \int d^dx \sqrt{\bs} \left[ d_i + \tfrac{1}{6} s \Rb \, d_i + \ldots \right]
\ee
where $d_i$ is shorthand for the dimension of the internal space and the dots denote curvature monomials build from higher orders of the spatial curvature tensors.

Substituting the expansion \eqref{heatkernel} into \eqref{trafo1} leads to a curvature expansion of the operator traces in terms of the intrinsic curvatures
\be\label{ftraces}
\begin{split}
{\rm tr}\left[ W(\Delta) \right] = & \, \frac{1}{(4 \pi)^{d/2}} \int d^dx \sqrt{\bs} \, \int_0^\infty ds \, s^{-d/2} \, \widetilde{W}(s) \, \left[ d_i + \tfrac{1}{6} \, s \, \Rb \, d_i  + \ldots \right] \\
= & \, \frac{1}{(4 \pi)^{d/2}} \int d^dx \sqrt{\bs} \left[ d_i \, Q_{d/2}[W]   + \tfrac{1}{6} \, d_i \, \Rb \, Q_{d/2-1}[W] + \ldots \right]
\end{split}
\ee
with the $Q$-functionals defined as
\be
Q_n[W] \equiv  \int_0^\infty ds \, s^{-n} \, \widetilde{W}(s) \, .
\ee
For $n > 0$, the definition of the $\Gamma$-functions allows to re-express 
these functionals as a Mellin transform
\be
Q_n[W] = \frac{1}{\Gamma(n)} \int_0^\infty dz \, z^{n-1} W(z) \, .
\ee

In order to write down the beta functions, it is useful to rewrite these $Q$-functionals in terms of dimensionless threshold functions. The $W$-functions appearing in the traces \eqref{tensortraces} and \eqref{scalartraces} are of the form
\be \label{eq:W}
W^{p,q}(w) \equiv \frac{\p_t R_k}{\left( P_k + w \right)^p \, \left(P_k\right)^q}  \, , \qquad
\widetilde{W}^{p,q}(w) \equiv \frac{R_k}{\left( P_k + w \right)^p \, \left(P_k\right)^q}  \, .
\ee
This functional form motivates defining
\be\label{treshold}
\begin{split}
\Phi^{p,q}_n(w) \equiv & \, \frac{1}{\Gamma(n)} \int_0^\infty \, dz z^{n-1} \, \frac{R^{(0)}(z) - z R^{(0)\prime}(z)}{\left( z + R^{(0)}(z) + w \right)^p \, \left(z + R^{(0)}(z) \right)^q}  \, , \\
\tilde{\Phi}^{p,q}_n(w) \equiv & \, \frac{1}{\Gamma(n)} \int_0^\infty \, dz z^{n-1} \, \frac{R^{(0)}(z)}{\left( z + R^{(0)}(z) + w \right)^p \, \left(z + R^{(0)}(z) \right)^q} \, ,
\end{split}
\ee
satisfying the recursion relations
\be\label{Phirec}
\frac{\p}{\p w} \Phi^{p,q}_n(w) = - p \, \Phi^{p+1,q}_n(w) \, , \qquad \frac{\p}{\p w} \tilde{\Phi}^{p,q}_n(w) = - p \, \tilde{\Phi}^{p+1,q}_n(w) \, . 
\ee
The definitions \eqref{treshold} then imply
\be
\begin{split}
Q_n\left[ W^{p,q} \right] = 2 k^{2(n-p-q+1)} \, \Phi^{p,q}_n(w/k^2) \, , \qquad
Q_n\left[ \widetilde{W}^{p,q} \right] = k^{2(n-p-q+1)} \, \tilde{\Phi}^{p,q}_n(w/k^2) \, . 
\end{split}
\ee
Inspecting the traces \eqref{tensortraces} and \eqref{scalartraces} one furthermore observes that the functions $W$ and $\widetilde{W}$ always appear in certain combinations. In this view it is also natural to introduce
\be\label{qfcts}
q^{p,q}_n(w) \equiv \Phi^{p,q}_n(w) - \half \eta_N \tilde{\Phi}^{p,q}_n(w)
\ee
where $\eta_N$ is the anomalous dimension of the Newton constant \eqref{Zk}.

Up to this point the profile function $R^{(0)}(z)$ entering the IR cutoff was left unspecified. For explicit computations it turns out to be convenient to work with the optimized cutoff \cite{opt}
\be\label{Ropt}
R^{(0)}(z) = \left(1-z\right) \theta\left( 1-z\right) \, .
\ee
For this choice the integrals appearing in the threshold functions \eqref{treshold} can be carried out analytically. The result degenerates in $q$ and reads
\be \label{eq:optThres}
\Phi^{p,q}_n(w) = \frac{1}{\Gamma(n+1)} \, \frac{1}{\left(1+w \right)^p} \, , \qquad \tilde{\Phi}^{p,q}_n(w) = \frac{1}{\Gamma(n+2)} \, \frac{1}{\left(1+w \right)^p} \, .
\ee
We will implement this choice of cutoff, when resumming the Matsubara modes in the next subsection.

\subsection{Resumming the Matsubara frequencies}
\label{App:resum}
Owed to the foliation structure, our flow equation naturally contains infinite sums over the Matsubara modes. For special choices of the regulator, as, e.g., the optimized cutoff \eqref{Ropt}, these sums can be carried out analytically. For the threshold functions appearing in the main text, this can be done via the master formula
\be \label{eq:mastersum}
\sum_{n = - \infty}^\infty \frac{1}{n^2 + x^2} =  \frac{\pi}{x \, \tanh(\pi x)} \, , \qquad
\sum_{n = - \infty}^\infty \frac{1}{n^2 - x^2} =  - \frac{\pi}{x \, \tan(\pi x)} 
\ee
related by analytic continuation. Based on this relation the sums in \eqref{eq:TgravTT1} and \eqref{Sgrav} can be carried out explicitly. Note that throughout this subsection we work exclusively with the optimized cutoff, such that the threshold functions take the form \eqref{eq:optThres}.

We start with the transverse-traceless sector, by defining
\be\label{Xifcts}
\Upsilon_{l}^{p,r} \equiv \sum_{n = - \infty}^\infty \left( \tfrac{1}{\epsilon} m^2 n^2 \right)^r \, \Phi_{l}^{p,q}(w_{\rm 2T}) \, , \qquad
\tilde{\Upsilon}_{l}^{p,r} \equiv \sum_{n = - \infty}^\infty \left( \tfrac{1}{\epsilon} m^2 n^2 \right)^r \, \tilde{\Phi}_{l}^{p,q}(w_{\rm 2T})
\ee
with $w_{\rm 2T}$ given in \eqref{def:w2T}. Inspecting \eqref{eq:TgravTT1}, we observe that the transverse-traceless sector contains functions $\Upsilon$ with $r=0$ only. For $p=1$ the sum is of the form \eqref{eq:mastersum} and yields
\be\label{Ups1}
\Upsilon_{l}^{1,0} = \frac{1}{\Gamma(l+1)} \frac{\pi}{m} \, \sqrt{\frac{\eps}{1-2\lambda}} \, \frac{1}{\tanh\left( \tfrac{\pi}{m} \sqrt{\epsilon \, (1-2\lambda)}  \right) } \, .
\ee
The $\Upsilon$ functions with higher values of $p$ are found by applying the recursion relation \eqref{Phirec}. Exploiting that $\Upsilon^{2,0}_l = \half \, \p_\lambda \, \Upsilon^{1,0}_l$, we obtain
\be\label{Ups2}
\Upsilon_{l}^{2,0} =  \, \tfrac{1}{\Gamma(l+1)} \tfrac{ \sqrt{\epsilon} \, \pi}{2m(1-2\lambda)} \left( 
\frac{1}{\sqrt{1-2\lambda}} \, \frac{1}{\tanh\left( \tfrac{\pi}{m} \sqrt{\epsilon \, (1-2\lambda)}  \right)} +  \frac{\sqrt{\epsilon} \pi}{m \, \sinh\left( \tfrac{\pi}{m} \sqrt{\epsilon \, (1-2\lambda)}  \right)^2}
\right) \, .
\ee

Resumming the threshold functions in the scalar sector is slightly more involved, but can by carried out along the same lines. Analogously to \eqref{Xifcts}, we start by introducing
\be\label{Psifcts}
\Psi_{l}^{p,r} \equiv \sum_{n = - \infty}^\infty \left( \tfrac{1}{\epsilon} m^2 n^2 \right)^r \, \Phi_{l}^{p,q}(w_0) \, , \qquad
\tilde{\Psi}_{l}^{p,r} \equiv \sum_{n = - \infty}^\infty \left( \tfrac{1}{\epsilon} m^2 n^2 \right)^r \, \tilde{\Phi}_{l}^{p,q}(w_0) \, .
\ee
The $w_0$, given in \eqref{def:w0}, leads to denominators which are (powers of) fourth order polynomials in $n$. This case can be reduced to the master formula \eqref{eq:mastersum} by first factorizing the denominators into second order polynomials and subsequently carrying out a partial fraction decomposition. For the purpose of computing the $\Psi$-functions recursively, it is useful to generalize the denominator $1+w_0$ by introducing auxiliary parameters $\beta, \gamma$
\be\label{genden}
1 + w_0 = \left. \gamma - 2 \lambda + \beta \, \tfrac{3d-4}{d-2} \,\tfrac{1}{\epsilon} m^2 n^2 - \tfrac{d-1}{d-2} \, \tfrac{1}{\lambda} \, m^4 n^4 \right|_{\beta = \gamma = 1} \, .
\ee 
The polynomial in $n$ factorizes into
\be \label{eq:factorization}
1 + w_0 =  -\Big(  \tfrac{d-1}{d-2} \tfrac{m^4}{\lambda} \Big) \Big(n^2 - w_+ \Big) \Big(n^2 - w_- \Big)
\ee
with
\be\label{wpm}
w_\pm =  \left. \frac{\eps \lambda}{m^2} \left( \frac{\beta (3d-4)}{2(d-1)} \pm \sqrt{ \left( \frac{\beta (3d-4)}{2(d-1)} \right)^2+  \frac{d-2}{d-1} \frac{1}{\lambda} (\gamma - 2 \lambda) } \right) \,
\right|_{\beta = \gamma = 1} \, .
\ee
Substituting \eqref{eq:factorization} into $\Psi^{1,0}_l$ and carrying out the partial fraction decomposition casts the sum into the form \eqref{eq:mastersum}, so that it is easily evaluated to be
\be\label{Psi1}
\Psi^{1,0}_l = \frac{1}{\Gamma(l+1)} \frac{d-2}{d-1} \frac{\pi \lambda }{m^4 \, (w_+ - w_-)} \left( \frac{1}{\sqrt{w_+} \, \tan(\pi \sqrt{w_+}) } - \frac{1}{\sqrt{w_-} \, \tan(\pi \sqrt{w_-})} \right) \, .
\ee

Based on this expression the other $\Psi^{p,r}_l$ can be found recursively by taking derivatives with respect to the auxiliary parameters introduced in \eqref{genden}. In particular the $\Psi^{p,0}_l$ satisfy
\be\label{reca}
 \Psi^{p+1,0}_l = - \left. \frac{1}{p} \, \frac{\p}{\p\gamma} \, \Psi^{p,0}_l \, \right|_{\beta = \gamma = 1} \, .
\ee
For $p=1$ this yields
\be\label{Psi2}
\begin{split}
\Psi^{2,0}_l = & \tfrac{1}{\Gamma(l+1)}\left( \tfrac{d-2}{d-1} \right)^2  \tfrac{\pi \lambda^2 }{m^8 \, (w_+ - w_-)^2} 
\bigg\{ \tfrac{2}{w_+ - w_-} \left( \tfrac{1}{\sqrt{w_+} \, \tan(\pi \sqrt{w_+}) } - \tfrac{1}{\sqrt{w_-} \, \tan(\pi \sqrt{w_-})}\right) \\ 
&  \, + \tfrac{1}{2 w_+} \left( \tfrac{1}{\sqrt{w_+} \, \tan(\pi \sqrt{w_+}) }  + \tfrac{\pi}{\sin^2(\pi \sqrt{w_+}) } \right) 
 + \tfrac{1}{2 w_-} \left( \tfrac{1}{\sqrt{w_-} \, \tan(\pi \sqrt{w_-}) } + \tfrac{\pi}{\sin^2(\pi \sqrt{w_-}) } \right) 
\bigg\} \, .
\end{split}
\ee
Furthermore, powers of $n^2$ in the numerator can be generated by taking derivatives with respect to $\beta$. For the $\Psi^{2,1}_l$ appearing in \eqref{Sgrav2} this implies
\be\label{recb}
\Psi^{2,1}_l = \left. - \frac{d-2}{3d-4} \, \frac{\p}{\p \beta} \, \Psi^{1,0}_l \, \right|_{\beta = \gamma = 1}
\ee
which evaluates to
\be\label{Psi21}
\begin{split}
\Psi^{2,1}_l = & \, \frac{1}{\Gamma(l+1)} \, \frac{(d-2)^2}{2(d-1)(3d-4)}
\frac{\pi \lambda}{m^4}  \frac{w_+ + w_-}{(w_+ - w_-)^2}\\ & \,  \bigg\{
\tfrac{1}{w_+ - w_-} \left( \tfrac{3w_+ +w_-}{\sqrt{w_+} \tan(\pi \sqrt{w_+})} 
-  \tfrac{3w_- +w_+}{\sqrt{w_-} \tan(\pi \sqrt{w_-})} \right) 
+   \tfrac{\pi}{\sin^2(\pi \sqrt{w_+})} +  \tfrac{\pi}{\sin^2(\pi \sqrt{w_-})} 
\bigg\} \, .
\end{split}
\ee
The corresponding expressions for the sums $\tilde{\Upsilon}$ and $\tilde{\Psi}$ are easily obtained from the results given above by taking into account the different prefactors \eqref{eq:optThres}, so that we refrain from giving explicit formulas.

For the purpose of writing the beta functions in a compact way, it is furthermore useful to introduce the analog of the $q$ functionals \eqref{qfcts} setting
\be\label{sumfcts}
T^{p,r}_l \equiv {\Upsilon}^{p,r}_l - \half \, \eta_N \, \tilde{\Upsilon}^{p,r}_l \, , \qquad S^{p,r}_l \equiv {\Psi}^{p,r}_l - \half \, \eta_N  \, \tilde{\Psi}^{p,r}_l
\ee
where the $T$ and $S$ are reminders that the threshold functions appear in the transverse-traceless and scalar traces, respectively. Given these results the evaluation of the functional traces \eqref{tensortraces} and \eqref{scalartraces} is now straightforward.

\section{Limits of the threshold functions}
\label{App:limit}
In order to understand the structure of the flow implied by the beta functions \eqref{eq:betaFct}, it is useful to study the two limiting cases where the time circle collapses, $T \rightarrow 0$, or becomes infinitely extended, $T\rightarrow \infty$. In Sect.\ \ref{sect.5} these two cases are called the compactification and decompactification limit respectively. This appendix then collects the corresponding limits of the threshold functions constructed in Sect.\ \ref{App:resum}. 

\subsection{The compactification limit $T \rightarrow 0$}
\label{App:C.1}
In the limit $T \rightarrow 0$, where the time circle collapses, the Matsubara mass \eqref{dimless} becomes infinite,
\be
\lim_{T \rightarrow 0} m = \infty \, . 
\ee
As a consequence the massive Matsubara modes decouple from the flow. The only term contributing to the Matsubara sums \eqref{Xifcts} and \eqref{Psifcts} is the zero mode $n=0$. It is then straightforward to verify that 
\be\label{Tlim1}
\begin{split}
\lim_{T \rightarrow 0} \Upsilon^{p,r}_l = \left\{ 
\begin{array}{ll}
\Phi^p_l(-2 \lambda)\, , & \qquad r = 0 \, , \\
0 \, , & \qquad r \not = 0 \le p-1 
\end{array}
\right.  
\end{split}
\ee
and 
\be\label{Slim1}
\begin{split}
\lim_{T \rightarrow 0} \Psi^{p,r}_l = \left\{ 
\begin{array}{ll}
\Phi^p_l(-2 \lambda)\, , & \qquad r = 0 \, , \\
0 \, , & \qquad r \not = 0 \le 2 p-1 \,\,\, .
\end{array}
\right. 
\end{split}
\ee
Taking the compactification limit thereby commutes with carrying out the Matsubara sums. In particular taking the limit $m \rightarrow \infty$ in eqs.\ \eqref{Ups1}, \eqref{Ups2}, \eqref{Psi1}, \eqref{Psi2} and \eqref{Psi21} confirms the general identities \eqref{Tlim1} and \eqref{Slim1}. This provides a non-trivial crosscheck for the threshold functions $\Upsilon$ and $\Psi$. As in the previous section, the result \eqref{Tlim1} and \eqref{Slim1} also holds for $\tilde \Upsilon$ and $\tilde \Psi$ if the r.h.s.\ is replaced by $\tilde{\Phi}^p_l(-2\lambda)$.

\subsection{The decompactification limit $T \rightarrow \infty$}
\label{App:C.2}
The limit $T  \rightarrow \infty$ is slightly more involved. In this case the Matsubara mass vanishes
\be\label{delim}
\lim_{T \rightarrow \infty} m = 0
\ee
so that the discrete sums essentially turn into a continuous integral. Inspecting the threshold functions one observes that this limit gives rise to finite results if and only if the threshold functions are of hyperbolic nature. Table \ref{tab:analyticStructure} then indicates that this is the case in two regions on the $g$-$\lambda$ plane
\be\label{cases}
\begin{array}{lcl}
{\rm I}:  & \qquad \lambda < -4 C_0 \, , \qquad & \qquad \epsilon = 1 \\
{\rm II}: & \qquad \lambda > 1/2 \, , \qquad & \qquad \epsilon = -1 \, .
\end{array}
\ee
Outside these regions the beta functions oscillate infinitely rapidly and are thus ill-defined. Therefore we will restrict ourselves to giving the limit behavior of the threshold functions appearing in the beta functions \eqref{eq:betaFct} for these two regions only. To simplify notation, it then turns out to be useful to extract the prefactors from \eqref{wpm} and define
\be\label{whpm}
\hat{w}_\pm =  \left. \frac{\beta (3d-4)}{2(d-1)} \pm \sqrt{ \left( \frac{\beta (3d-4)}{2(d-1)} \right)^2+  \frac{d-2}{d-1} \frac{1}{\lambda} (\gamma - 2 \lambda) } \;
\right|_{\beta = \gamma = 1}
\ee
which are manifestly positive in both regions.

Taking the limit $m \rightarrow 0$ in the tensor sector, \eqref{Ups1} and \eqref{Ups2}, is straightforward and yields
\be
\lim_{m \rightarrow 0} \, m \, \Upsilon^{p,0}_l = \frac{\epsilon \, \pi}{p \, \Gamma(l+1)} \, \frac{1}{(1-2\lambda)^{p-1}} \, \frac{1}{|1-2\lambda|^{1/2}} \, , \qquad p = 1,2 \, . 
\ee

The scalar sector is slightly more involved. Here we first note that since $\epsilon = 1, \lambda < 0$ in case I and $\epsilon = -1, \lambda > 0$ in case II, the arguments of the square-roots appearing in \eqref{Psi1}, \eqref{Psi2} and \eqref{Psi21} are actually negative which requires the analytic continuation of the formulas to hyperbolic functions. Subsequently taking the limit \eqref{delim} in \eqref{Psi1} gives
\be\label{Psi1lim}
\lim_{m \rightarrow 0} \, m \, \Psi^{1,0}_l = - \frac{\epsilon \, \pi}{\Gamma(l+1)} \, \frac{d-2}{d-1} \, \frac{1}{|\lambda|^{1/2}} \, \frac{1}{\hat{w}_+ - \hat{w}_-} \left( \frac{1}{\sqrt{\hat{w}_+}} - \frac{1}{\sqrt{\hat{w}_-}} \right) \, .   
\ee
The higher order threshold functions can then again be obtained via the recursion relations \eqref{reca} and  \eqref{recb}. In particular
\be\label{Psi2lim}
\begin{split}
\lim_{m \rightarrow 0} \, m \, \Psi^{2,0}_l =  & \, \tfrac{\epsilon \, \pi}{\Gamma(l+1)} \, \left( \tfrac{d-2}{d-1} \right)^2 \, \tfrac{1}{2 \, |\lambda|^{3/2}} \, \tfrac{1}{(\hat{w}_+ - \hat{w}_-)^2} 
\left\{ \tfrac{4}{\hat{w}_+ - \hat{w}_-} \left( \tfrac{1}{\sqrt{\hat{w}_+}} - \tfrac{1}{\sqrt{\hat{w}_-}} \right) + \tfrac{1}{\sqrt{\hat{w}_+}^3} + \tfrac{1}{\sqrt{\hat{w}_-}^3} \right\} \, , \\
\lim_{m \rightarrow 0} \, m \, \Psi^{2,1}_l = & \, - \tfrac{\epsilon \, \pi}{\Gamma(l+1)} \, \tfrac{(d-2)^2}{d-1}  \, \tfrac{1}{3d-4} \, \tfrac{1}{2 \, |\lambda|^{1/2}} \, \tfrac{\hat{w}_+ + \hat{w}_-}{(\hat{w}_+ - \hat{w}_-)^3} 
\, \left\{  \tfrac{3 \hat{w}_+ + \hat{w}_-}{\sqrt{\hat{w}_+}} - \tfrac{3 \hat{w}_- +  \hat{w}_+}{\sqrt{\hat{w}_-}}  \right\} \, . 
\end{split}
\ee
Again the corresponding expressions for the sums $\tilde{\Upsilon}$ and $\tilde{\Psi}$ are easily obtained from the results given above by taking into account the different prefactors \eqref{eq:optThres}, so we refrain from giving explicit expressions. 

\end{appendix}
\bibliographystyle{unsrt}

\end{document}